\documentclass[letterpaper,11pt]{article}
\pdfoutput=1

\usepackage{jheppub}
\usepackage{multirow}

\usepackage{subfig}
\usepackage{xspace}
\usepackage[countmax]{subfloat}

\usepackage{amssymb}
\usepackage{amsmath}
\usepackage{cancel}
\usepackage{color}
\usepackage{braket}
\usepackage{graphicx}
\usepackage{multirow}
\usepackage{verbatim}
\usepackage{amsthm}
\usepackage{slashed}
\usepackage{wasysym}
\usepackage{simplewick}
\usepackage{mathtools}
\usepackage{soul}
\usepackage{xspace}

\definecolor{dancomment}{RGB}{0,159,0}

\def\cB{\mathcal{B}}

\def\cL{\mathcal{L}}

\def\cO{\mathcal{O}}

\def\cP{\mathcal{P}}

\def\cW{\mathcal{W}}

\def\cY{\mathcal{Y}}

\def\tr{{\rm tr}}

\def\nn{{\nonumber}}

\newcommand{\hard}{\mathrm{hard}}
\newcommand{\dyn}{\mathrm{dyn}}

\newcommand{\BPS}{\mathrm{BPS}}

\newcommand{\Eq}[1]{Equation~\eqref{#1}}

\DeclareRobustCommand{\Sec}[1]{Sec.~\ref{#1}}

\DeclareRobustCommand{\Tab}[1]{Table~\ref{#1}}

\DeclareRobustCommand{\Fig}[1]{Fig.~\ref{#1}}

\DeclareRobustCommand{\Eq}[1]{Eq.~(\ref{#1})}
\DeclareRobustCommand{\Eqs}[2]{Eqs.~(\ref{#1}) and (\ref{#2})}

\def\be{\begin{equation}}
\def\ee{\end{equation}}

\newcommand{\SCETi}{\mbox{${\rm SCET}_{\rm I}$}\xspace}
\newcommand{\SCETii}{\mbox{${\rm SCET}_{\rm II}$}\xspace}

\def\l{\langle}
\def\r{\rangle}

\def\bt{\beta}

\newcommand{\Sl}[1]{\slashed{#1}}

\allowdisplaybreaks[3]

\newcommand{\eq}[1]{Eq.~\eqref{eq:#1}}
\newcommand{\eqs}[2]{Eqs.~\eqref{eq:#1} and \eqref{eq:#2}}
\renewcommand{\sec}[1]{Sec.~\ref{sec:#1}}

\newcommand{\fig}[1]{Fig.~\ref{fig:#1}}

\newcommand{\ord}[1]{\mathcal{O}(#1)}

\newcommand{\mae}[3]{\langle#1\rvert#2\rvert#3\rangle}
\newcommand{\Mae}[3]{\bigl\langle#1\bigr\rvert#2\bigr\rvert#3\bigr\rangle}

\newcommand{\df}{\mathrm{d}}
\newcommand{\img}{\mathrm{i}}

\newcommand{\sdt}{\!\cdot\!}

\newcommand{\al}{\alpha}

\newcommand\bn{{\bar n}}
\newcommand{\ga}{\gamma}

\newcommand{\ve}{\varepsilon}
\newcommand{\la}{\lambda}

\newcommand{\w}{\omega}
\newcommand{\balpha}{{\bar \alpha}}
\newcommand{\bbeta}{{\bar \beta}}

\newcommand{\vT}{\bar{T}}

\newcommand{\vC}{\vec{C}}
\newcommand{\vO}{\vec O}

\newcommand{\lp}{\tilde p}        
\newcommand{\ldel}{\tilde \delta} 

\newcommand{\bnP}{\overline {\mathcal P}}

  \newcount\hour \newcount\minute
  \hour=\time \divide \hour by 60 \minute=\time
  \newcommand{\todaytime}{\today \ -- \number\hour :\ifnum \minute<10 0\fi\number\minute}

\preprint{MIT--CTP 4748}

\title{Building Blocks for Subleading Helicity Operators}

\author{Daniel W. Kolodrubetz, Ian Moult, Iain W. Stewart}

\affiliation{Center for Theoretical Physics, Massachusetts Institute of Technology, Cambridge, MA 02139, USA}

\emailAdd{dkolodru@mit.edu}
\emailAdd{ianmoult@mit.edu}
\emailAdd{iains@mit.edu}

\abstract{On-shell helicity methods provide powerful tools for determining scattering amplitudes, which have a one-to-one correspondence with leading power helicity operators in the Soft-Collinear Effective Theory (SCET) away from singular regions of phase space. We show that helicity based operators are also useful for enumerating power suppressed SCET operators, which encode subleading amplitude information about singular limits. In particular, we present a complete set of scalar helicity building blocks that are valid for constructing operators at any order in the SCET power expansion.  We also describe an interesting angular momentum selection rule that restricts how these building blocks can be assembled. 
}


\begin{document} 

\maketitle

\section{Introduction}\label{sec:intro}

The use of on-shell helicity amplitudes has proved fruitful for the study of scattering amplitudes in gauge theories and gravity (see e.g. \cite{Dixon:1996wi,Elvang:2013cua,Dixon:2013uaa,Henn:2014yza} for pedagogical reviews). By using external states of definite helicity, gauge redundancies are removed and the underlying symmetries of the theory are made manifest. Helicity based techniques have also proved to be a powerful organizing principle for studying operator bases in effective field theories. Recently this has been demonstrated by the use of helicity arguments \cite{Cheung:2015aba} to determine the pattern of non-renormalization for dimension 6 operators in the Standard Model effective theory \cite{Alonso:2014rga}, as well as for constructing hard scattering operator bases for collider processes~\cite{Moult:2015aoa} in the Soft Collinear Effective Theory (SCET)~\cite{Bauer:2000ew, Bauer:2000yr, Bauer:2001ct, Bauer:2001yt}.

\begin{figure}[t!]
%
%
\begin{center}
\includegraphics[width=0.23\columnwidth]{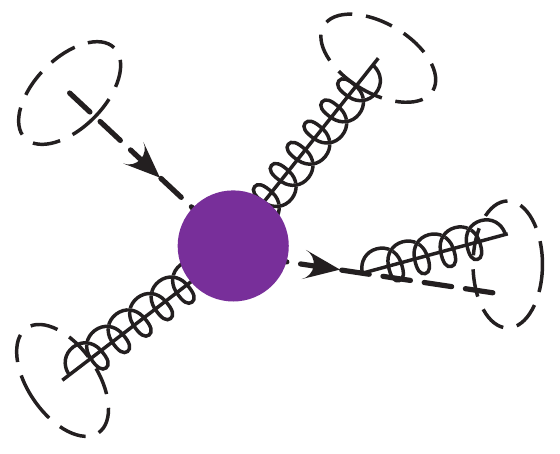} 
\hspace{0.1cm}
\includegraphics[width=0.23\columnwidth]{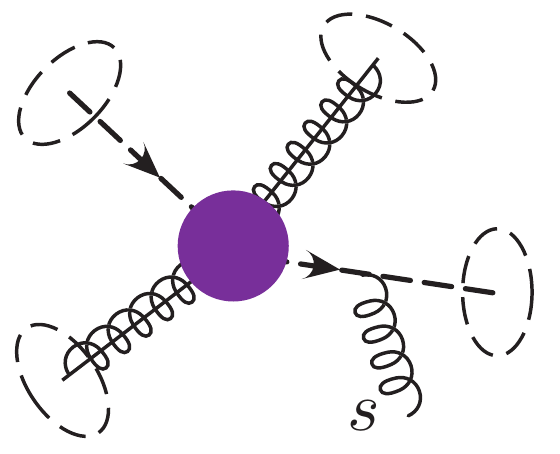} 
\hspace{0.1cm}
\includegraphics[width=0.23\columnwidth]{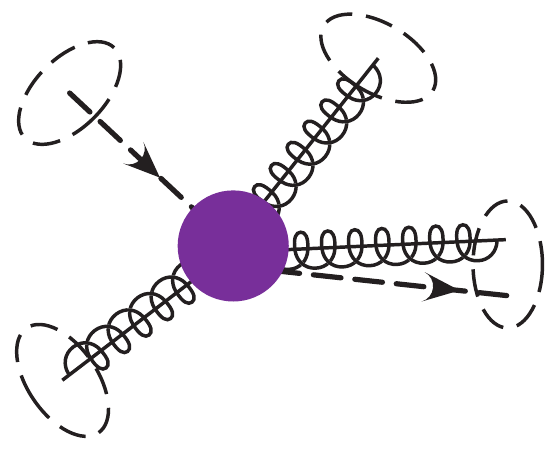} 
\hspace{0.1cm}
\includegraphics[width=0.23\columnwidth]{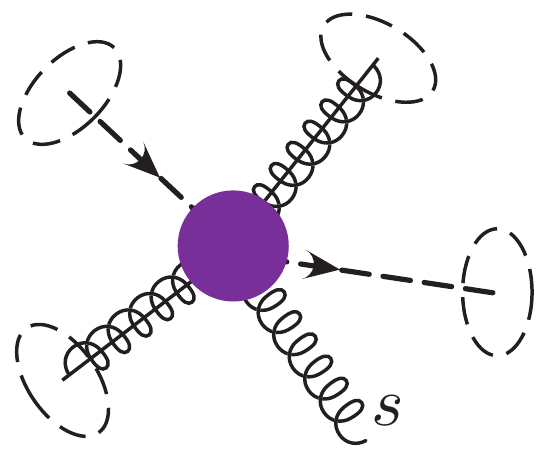} 
\raisebox{0cm}{ \hspace{-0.2cm} 
  $a$)\hspace{3.4cm}
  $b$)\hspace{3.4cm} 
  $c$)\hspace{3.4cm}
  $d$)\hspace{4cm} } 
\\[-25pt]
\end{center}
\vspace{-0.4cm}
\caption{ 
Example of scattering amplitudes with energetic particles in four distinct regions of phase space, at leading power in a) and b), and subleading power in c) and d).  There is an extra collinear gluon in a) from splitting, and in b) there is an extra gluon from soft emission. In c) the extra energetic gluon is collinear with the quark, but occurs without a nearly onshell parent propagator. Likewise in d) the extra soft emission amplitude is subleading. } 
\label{fig:subleadingamp}
\end{figure}

Effective field theories provide an important tool for studying gauge theories, where simplified or universal behavior often appears in specific limits. They allow for a systematic expansion that enables questions about subleading corrections to be rigorously studied. Of particular interest, both theoretically and phenomenologically, are the soft and collinear limits of gauge theories. The behavior of amplitudes \cite{Weinberg:1965nx} and cross sections in the soft and collinear limits, and the factorization theorems~\cite{Collins:1985ue,Collins:1989gx,Collins:1988ig,Collins:1984kg} describing their behavior in these limits, have primarily been studied at leading power in the expansion. The leading soft and collinear limits give rise to the leading singular behavior of collider observables.  Examples include the $1/\tau$ terms for thrust \cite{Catani:1991kz,Catani:1992ua}, which dominate in the $\tau\to 0$ limit, or the $1/(1-z)$ terms for threshold resummation~\cite{Sterman:1986aj,Catani:1989ne}, which dominate in the limit $z\to 1$ (here $z=Q^2/\hat s$, with $Q^2$ the invariant mass of the final state and $\hat s$ the center-of-mass energy). An understanding of the subleading soft and collinear limits is also of considerable interest, both at the amplitude level, for understanding the subleading behavior of gauge theory and gravity amplitudes \cite{Low:1958sn,Burnett:1967km,DelDuca:1990gz,Casali:2014xpa,Schwab:2014xua,Cachazo:2014fwa,Larkoski:2014hta,He:2014bga,Bern:2014oka,Broedel:2014fsa,Zlotnikov:2014sva,Lysov:2014csa,White:2014qia,Larkoski:2014bxa,Kapec:2015ena,Strominger:2015bla}, and at the cross section level \cite{Lee:2004ja,Mannel:2004as,Bosch:2004cb,Beneke:2004in,Laenen:2008gt,Laenen:2008ux,Grunberg:2009yi,Laenen:2010uz,Almasy:2010wn,Freedman:2013vya,Freedman:2014uta,Bonocore:2014wua,deFlorian:2014vta,Bonocore:2015esa}, where they determine the structure of the ${\cal O}(\tau^0)$ corrections for thrust and the ${\cal O}((1-z)^0)$ corrections in the threshold expansion, and allow questions about the universality of these terms to be addressed.  An example of the type of amplitudes that are described at leading and subleading power are shown in \fig{subleadingamp}. For leading power amplitudes with an extra collinear or soft gluon emission, such as those in \fig{subleadingamp}a,b, the extra gluon is accompanied by the enhancement from an additional nearly onshell propagator. In contrast, in the subleading amplitudes in \fig{subleadingamp}c,d we have an extra gluon emission without this enhancement. 

%
%

SCET is an effective field theory describing the dynamics of collinear and soft particles in the presence of a hard scattering interaction with a systematic expansion in a power counting parameter $\lambda\ll 1$.  It can be used to study both the leading and subleading corrections in soft and collinear limits, and several SCET analyses have been performed at subleading power~\cite{Beneke:2002ni,Chay:2002vy,Manohar:2002fd,Pirjol:2002km,Beneke:2002ph,Bauer:2003mga,Lee:2004ja,Mannel:2004as,Bosch:2004cb,Beneke:2004in,Arnesen:2006vb,Arnesen:2006dc,Lee:2006nr,Benzke:2010js,Mateu:2012nk,Freedman:2013vya,Freedman:2014uta,Larkoski:2014bxa}.  Collinear modes in the effective field theory are expanded about the lightlike direction of jets, shown as dashed circles in \fig{subleadingamp}, and the fields describing these modes carry a lightlike reference vector with respect to which helicities can be naturally defined. Instead of considering operators formed from Lorentz and Dirac structures, each of which contributes to multiple states with different helicity combinations, one can use helicity fields associated with external states of definite helicity with respect to the jet axes~\cite{Moult:2015aoa}. Using helicity based building blocks to construct operators greatly simplifies finding a minimal operator basis for processes with many active partons, and facilitates the matching to fixed order calculations which are often performed using spinor helicity techniques \cite{Dixon:1996wi,Elvang:2013cua,Dixon:2013uaa,Henn:2014yza} .


In this paper, we show that helicity operators also greatly facilitate the study of subleading power corrections in SCET. We develop a complete set of collinear and soft gauge invariant helicity building blocks, valid for constructing operators at any order in the power expansion. The use of these helicity building blocks greatly simplifies the construction of a complete subleading power operator basis in the effective theory, and makes various symmetries manifest. Additionally, it eliminates the need to consider equation of motion relations to remove redundant operators. The subleading helicity operators obey interesting (and simple) angular momentum selection rules, which we discuss. In a companion paper~\cite{subhel:long}, we will provide a more detailed discussion of various aspects of these subleading helicity operators, including the construction of a complete basis of operators for processes involving two collinear directions to ${\cal O}(\lambda)$ and ${\cal O}(\lambda^2)$ in the power expansion.

Below in \sec{scethelicity} we review salient features and notation from SCET with and without helicity operators. In \sec{helops} we derive the complete set of helicity building blocks that are required for constructing operators at any order in the SCET power expansion. We carefully treat both collinear and soft degrees of freedom, and describe how the helicity basis is also convenient for organizing color degrees of freedom, including the soft Wilson lines arising from eikonalized particles participating in the hard scattering. In \sec{ang_cons} we discuss angular momentum selection rules which play an important role at subleading power when multiple collinear fields are present in the same collinear sector. These rules can significantly reduce the number of operators in the basis for a given process. In \sec{example} we demonstrate the utility of the helicity building blocks by constructing an operator basis involving two collinear quark fields, and two collinear gluon fields with two hard scattering directions (relevant for applications to Drell-Yan, $e^+e^-\to$ dijets, or DIS). We conclude in \sec{conclusions}.

\section{SCET and Helicity Fields}\label{sec:scethelicity}

SCET is an effective field theory describing the dynamics of collinear and soft particles in the presence of a hard interaction~\cite{Bauer:2000ew, Bauer:2000yr, Bauer:2001ct, Bauer:2001yt, Bauer:2002nz}. The collinear particles are energetic and collimated along jet directions, while the soft particles describe low energy radiation emitted from the jets. 
We employ two light-like reference vectors for each collinear direction, $n_i^\mu$ and $\bn_i^\mu$ such that $n_i^2 = \bn_i^2 = 0$ and $n_i\sdt\bn_i = 2$. A typical choice is $n_i^\mu = (1, \vec{n}_i)$, $\bn_i^\mu = (1, -\vec{n}_i)$
where $\vec{n}_i$ is a unit three-vector. Given a $n_i^\mu$ and $\bn_i^\mu$, any four-momentum $p$ can be decomposed as
\begin{equation} \label{eq:lightcone_dec}
p^\mu = \bn_i\sdt p\,\frac{n_i^\mu}{2} + n_i\sdt p\,\frac{\bn_i^\mu}{2} + p^\mu_{n_i\perp}\
\,.\end{equation}
An ``$n_i$-collinear'' quark or gluon has momentum $p^\mu$ close to the $\vec{n}_i$ direction, so that the components $(n_i\cdot p, \bn_i \cdot p, p_{n_i\perp}) \sim \,Q(\la^2,1,\la)$, where $Q$ is the scale of the hard scattering.  Here $\la \ll 1$ is a small parameter determined by the form of the measurement or kinematic restrictions under consideration. Soft particles have a homogeneously small scaling for their momentum components, which is typically given by $p^\mu\sim \lambda^2$ (termed ultrasoft) or ${p^\mu\sim \lambda}$ (termed soft), again depending on the type of measurement. For convenience we will predominantly concern ourselves with \SCETi where the dynamics is dominated by collinear and ultrasoft particles. To ensure that two different directions $n_i$ and $n_j$ refer to distinct collinear sectors, they have to be well separated, meaning $n_i \cdot n_j  \gg \la^2$ for $i\neq j$ ~\cite{Bauer:2002nz}. Two different reference vectors, $n_i$ and $n_i'$, with $n_i\cdot n_i' \sim \ord{\lambda^2}$ both describe the same jet and corresponding collinear physics. Thus, each collinear sector can be labelled by any member of a set of equivalent vectors, $\{n_i\}$. This freedom is manifest as a symmetry of the effective theory known as reparametrization invariance (RPI) \cite{Manohar:2002fd,Chay:2002vy}.

SCET is formulated as an expansion in powers of $\la$, and has manifest power counting at all stages of a calculation. A momentum space multipole expansion is used to construct the effective theory, and is carried out by expanding momenta into label and residual components with respect to the reference vector
\begin{equation} \label{eq:label_dec}
p^\mu = \lp^\mu + k^\mu = \bn_i \sdt\lp\, \frac{n_i^\mu}{2} + \lp_{n_i\perp}^\mu + k^\mu\,.
\,\end{equation}
Here, $\bn_i \cdot\lp \sim \la^0$ and $\lp_{n_i\perp} \sim \la$ are the large label momentum components, while $k^\mu \sim \la^2 $ is a smaller residual momentum. The full theory quark and gluon fields are expanded to obtain fields with momenta of definite scaling, namely collinear quark and gluon fields for each collinear direction, as well as ultrasoft quark and gluon fields. Independent ultrasoft and collinear gauge symmetries are enforced on the theory, and enable the distinction between collinear and ultrasoft gluon modes~\cite{Bauer:2001yt}.

The SCET fields for $n_i$-collinear quarks and gluons, $\xi_{n_i,\lp}(x)$ and $A_{n_i,\lp}(x)$, are labeled by their collinear direction $n_i$ and their large momentum $\lp$. They are typically written in position space with respect to the residual momentum and in momentum space with respect to the large momentum components. The large label momentum is obtained from the label momentum operator $\cP_{n_i}^\mu$, e.g. $\cP_{n_i}^\mu\, \xi_{n_i} = \lp^\mu\, \xi_{n_i}$~\cite{Bauer:2001ct}. For later convenience, we define $\bnP_{n_i} = \bn_i \sdt\cP_{n_i}$, which picks out the large momentum component. Derivatives acting on the fields pick out the residual momentum dependence, $i \partial^\mu \sim k^\mu \sim \la^2 Q$. The ultrasoft degrees of freedom in the effective theory are described by fields $q_{us}(x)$ and $A_{us}(x)$ without label momenta. They are able to exchange residual momenta between the jets in different collinear sectors.

The SCET Lagrangian is expanded as a power series in $\lambda$
\begin{align} \label{eq:SCETLagExpand}
\cL_{\text{SCET}}=\cL_\dyn + \cL_\hard
   = +\sum_{i\geq0} \cL^{(i)}+ \sum_{i\geq0} \cL_\hard^{(i)} \,,
\end{align}
where the superscript $(i)$ denotes objects at ${\cal O}(\lambda^i)$ in the power counting.
Here the ${\cal L}^{(i)}$ describe the interactions of ultrasoft and collinear paraticles within the effective theory, with the dynamics being dominated by the leading power Lagrangian ${\cal L}^{(0)}$. Expressions for the leading power Lagrangian can be found in~\cite{Bauer:2001yt}, and expressions for ${\cal L}^{(1)}$, and ${\cal L}^{(2)}$ can be found in~\cite{Bauer:2003mga} (see also~\cite{Pirjol:2002km,Beneke:2002ni,Chay:2002vy,Manohar:2002fd,Beneke:2002ph}).  Particles that exchange large momentum of $\ord{Q}$ between different jets are off-shell by $\ord{n_i\cdot n_j Q^2}$. These are integrated out by matching QCD onto SCET to give hard scattering operators $O^{(i)}$ that appear in $\cL_\hard^{(i)}$. The hard scattering operators are formed from collinear and ultrasoft gauge invariant products of collinear and ultrasoft fields, along with derivative operators and Wilson lines. It is convenient to work with a minimal set of collinear gauge invariant operators, which are referred to as collinear building blocks.  Using the equations of motion and Wilson line identities, it can be shown that a complete set of collinear and ultrasoft building blocks for the \SCETi hard scattering operators $O^{(i)}$ at any order in the power counting are given by~\cite{Marcantonini:2008qn}:
\begin{align} \label{eq:PC}
\begin{tabular}{| l | c | c |c |c|c| r| }
  \hline                       
  Operator & $\cB_{n_i\perp}^\mu$ & $\chi_{n_i}$& $\cP_{\!\perp}^\mu$&$q_{us}$&$D_{us}^\mu$ \\
  Power Counting & $\lambda$ &  $\lambda$& $\lambda$& $\lambda^3$& $\lambda^2$ \\
  \hline  
\end{tabular}
\end{align}
Here the ultrasoft quark field $q_{us}$ and covariant derivative $iD_{us}^\mu=i\partial^\mu+gA_{us}^\mu$ are the same as in a standard gauge theory. The collinear gauge invariant building blocks for collinear quarks/antiquarks and gluons, each with two spin states, are defined as
\begin{align} \label{eq:chiBbare}
\chi_{{n_i}}(x) &=  W_{n_i}^\dagger(x)\, \xi_{n_i}(x)
\,,
& \cB_{{n_i}\perp}^\mu(x)
&= \frac{1}{g}\,\bigl[ W_{n_i}^\dagger(x)\,\img D_{{n_i}\perp}^\mu W_{n_i}(x)\bigr]
 \,. 
\end{align}
Here the derivative $i D_{{n_i}\perp}^\mu = \cP^\mu_{{n_i}\perp} + g A^\mu_{{n_i}\perp}$ acts only within the square brackets. To ensure uniform power counting we decompose derivatives acting on an $n_i$-collinear field in terms of the $n_i$, $\bn_i$ basis, so the fact that $\perp$ means perpendicular to $n_i$ and $\bn_i$ is always clear from the context and we can write $\cP^\mu_{{n_i}\perp}$ as $\cP^\mu_{\perp}$ . The collinear Wilson lines appearing in \Eq{eq:chiBbare} are defined as
\begin{equation} \label{eq:Wn}
W_{n_i}(x) = \biggl[~\sum_\text{perms} \exp\Bigl(-\frac{g}{\bnP_{n_i}}\,\bn\sdt A_{n_i}(x)\Bigr)~\biggr]\,.
\end{equation}
Only the ${\cal P}_\perp^\mu$ derivative is needed in \eq{PC} since $i n_i\cdot\partial$ can be eliminated with the equations of motion. The power counting given in \eq{PC} is determined by demanding that the leading power action for the SCET fields is ${\cal O}(\lambda^0)$. The power counting for a composite operator is obtained by adding up the powers for the building blocks it contains. When building hard scattering operators it is often convenient to specify the ${\cal O}(\lambda^0)$ momentum of the collinear building blocks, via a $\omega$ momentum label $\chi_{n_i,\w} = \bigl[\delta(\w - \bnP_{n_i}) \chi_{n_i}\bigr]$ and $\cB_{{n_i}\perp,\w}^\mu = \bigl[  \delta(\w + \bnP_{n_i}) \cB_{\perp,\w}^\mu\bigr]$.  

Since the building blocks in \eq{PC} carry vector or spinor Lorentz indices they must be contracted to form scalar operators, which involves the use of objects like $\{n_i^\mu, \bn_i^\mu, \gamma^\mu, g^{\mu\nu},$ $\epsilon^{\mu\nu\sigma\tau}\}$.  For operators describing many jet directions or for operators at subleading power, constructing a minimal basis in this manner becomes difficult. Rather than dealing with contractions of vector and spinor indices, one can exploit a decomposition into operators with definite helicity, and work with building blocks that are scalars.\footnote{Generically when we say scalar building blocks, we are not accounting for their transformations under parity. Constraints from parity transformations are easy to include, see~\cite{Moult:2015aoa}.} For SCET operators this approach was formalized in~\cite{Moult:2015aoa} by defining helicity building block fields for the construction of leading power operators for jet processes. It takes advantage of the fact that collinear SCET fields are themselves collinear gauge invariant, and are each associated with a fixed external label direction with respect to which helicities can naturally be defined. We will follow the notation and conventions of~\cite{Moult:2015aoa}. We first define  collinear gluon and quark fields of definite helicity as
\begin{subequations}
	\label{eq:cBpm_quarkhel_def}
\begin{align} 
\label{eq:cBpm_def}
\cB^a_{i\pm} &= -\ve_{\mp\mu}(n_i, \bn_i)\,\cB^{a\mu}_{n_i\perp,\w_i}
\,, \\
\label{eq:quarkhel_def}
 \chi_{i \pm}^\alpha &= \frac{1\,\pm\, \gamma_5}{2} \chi_{n_i, - \omega_i}^\alpha
\,,\qquad\quad
\bar{\chi}_{i \pm}^\balpha =  \bar{\chi}_{n_i, - \omega_i}^\balpha \frac{1\,\mp\, \gamma_5}{2}\,,
\end{align}
\end{subequations}
where $a$, $\alpha$, and $\balpha$ are adjoint, $3$, and $\bar 3$ color indices respectively, and by convention the $\omega_i$ labels on both the gluon and quark building block are taken to be outgoing.  Using the standard spinor helicity notation (see e.g. \cite{Dixon:1996wi} for an introduction) we have
\begin{align} \label{eq:braket_def}
|p\rangle\equiv \ket{p+} &= P_R\, u(p)
  \,,
 & |p] & \equiv \ket{p-} = P_L \, u(p)
  \,, \\
\bra{p} \equiv \bra{p-} &= \mathrm{sgn}(p^0)\, \bar{u}(p)\, P_R
  \,, 
 & [p| & \equiv \bra{p+} = \mathrm{sgn}(p^0)\, \bar{u}(p)\, P_L
  \,, \nn 
\end{align}
with $p$ lightlike, $P_L=(1-\gamma_5)/2$ and $P_R=(1+\gamma_5)/2$. The polarization vector of an outgoing gluon with momentum $p$ can be written
\begin{equation}
 \ve_+^\mu(p,k) = \frac{\mae{p+}{\ga^\mu}{k+}}{\sqrt{2} \langle kp \rangle}
\,,\qquad
 \ve_-^\mu(p,k) = - \frac{\mae{p-}{\ga^\mu}{k-}}{\sqrt{2} [kp]}
\,,\end{equation}
where $k\neq p$ is an arbitrary light-like reference vector, chosen to be $\bn_i$ in \eq{cBpm_def}.  The lowest order Feynman rules for these fields are simple. For example, for an outgoing gluon with polarization $\pm$, momentum $p$ ($p^0>0$), and color $a$ we have ${\l g_\pm^a(p) | \cB_{i \pm}^b |0 \r = \delta^{ab} \ldel (\lp_i -p)}$, while for an incoming quark ($p^0<0$) with helicity $\pm$ and color $\alpha$ we have
$\Mae{0}{\chi^\beta_{i\pm}}{q_\pm^\balpha(-p)} = \delta^{\beta\balpha}\, \ldel(\lp_i - p)\, \ket{(-p_i)\pm}_{n_i}$.   Here we define the spinors with an SCET projection operator by $|p \pm \r_{n_i}\equiv \frac{\slashed{n}_i \slashed{\bar{n}}_i}{4} | p \pm \r$ and the $\tilde\delta(\tilde p_i-p)$ indicate that the momentum label in the building block field matches that of the state. The full set of Feynman rules are given in~\cite{Moult:2015aoa}.


To take advantage of the fact that fermions always come in pairs, Ref.~\cite{Moult:2015aoa} defined the currents
\begin{align} \label{eq:jpm_def}
 J_{ij\pm}^{\balpha\beta}
& = \mp\, \sqrt{\frac{2}{\omega_i\, \omega_j}}\, \frac{   \ve_\mp^\mu(n_i, n_j) }{\langle n_j\mp | n_i\pm\rangle}   \, \bar{\chi}^\balpha_{i\pm}\, \gamma_\mu \chi^\beta_{j\pm}
\,, \\
 J_{ij0}^{\balpha\beta}
& =\frac{2}{\sqrt{\vphantom{2} \omega_i \,\omega_j}\,  [n_i n_j] } \bar \chi^\balpha_{i+}\chi^\beta_{j-}
\,, \qquad
(J^\dagger)_{ij0}^{\balpha\beta}=\frac{2}{\sqrt{ \vphantom{2} \omega_i \, \omega_j}  \langle n_i  n_j \rangle  } \bar \chi^\balpha_{i-}\chi^\beta_{j+}
. \nn
\end{align}
These currents are manifestly invariant under the RPI-III symmetry of SCET, which takes $n_i^\mu \to e^\alpha n_i^\mu$ and $\bn_i^\mu \to e^\alpha \bn_i^\mu$, since $\omega_i\sim\bn_i$ and the $| n_i \rangle\sim\sqrt{n_i}$.  In general these currents consist of two spin-$1/2$ objects whose spin quantum numbers are specified along different axes, $\hat n_i$ and $\hat n_j$.  If we consider back-to-back collinear directions $n$ and $\bn$, then the two axes are the same, and these currents have definite helicity, given by
 \begin{align} \label{eq:jpm_back_to_bacjdef}
 & h=\pm 1:
 & J_{n \bn \pm}^{\balpha\beta}
 & = \mp\, \sqrt{\frac{2}{\omega_n\, \omega_\bn}}\, \frac{   \ve_\mp^\mu(n, \bn) }{\langle \bn \mp | n \pm\rangle}   \, \bar{\chi}^\balpha_{n\pm}\, \gamma_\mu \chi^\beta_{\bn \pm}
 \,, \\
 & h=0:
 & J_{n \bn 0}^{\balpha\beta}
 & =\frac{2}{\sqrt{\vphantom{2} \omega_n \,\omega_\bn}\,  [n \bn] } \bar \chi^\balpha_{n+}\chi^\beta_{\bn-}
 \,, \qquad
 (J^\dagger)_{n \bn 0}^{\balpha\beta}=\frac{2}{\sqrt{ \vphantom{2} \omega_n \, \omega_\bn}  \langle n  \bn \rangle  } \bar \chi^\balpha_{n-}\chi^\beta_{\bn+}
 . \nn
 \end{align}
The currents $J_{n \bn \pm}^{\balpha \beta}$ have helicity $h=\pm1$ along $\hat n$ respectively. The current $J_{n \bn 0}^{\balpha \beta} + (J^\dagger)_{n \bn 0}^{\balpha \beta}$ transforms as a scalar under rotations about the $n$ axis, i.e. has helicity zero (while the current $J_{n \bn 0}^{\balpha \beta} - (J^\dagger)_{n \bn 0}^{\balpha \beta}$ transforms as a pseudoscalar). We choose to use the $0$ subscript in both the back-to-back and non-back-to-back cases, to emphasize the helicity for the former case and conform with our notation for subleading currents below.

Together, the gluon building blocks $\cB^a_{i\pm}$ and the current building blocks $J_{ij\pm}^{\bar\alpha\beta}$, $J_{ij\,0}^{\bar\alpha\beta}$, and $(J^\dagger)_{ij\, 0}^{\bar\alpha\beta}$ suffice for the construction of leading power operators for all hard processes. (The only exceptions are hard processes that start at a power suppressed order.) All these objects behave like scalars under the Lorentz group, and can trivially be combined to form hard scattering operators by simple multiplication. The construction of leading power operators of this type was the focus of~\cite{Moult:2015aoa}.  We review below the organization of color structures in the leading power hard scattering operators and the decoupling of soft and collinear degrees of freedom using the BPS field redefinition. Then, in the next section we will extend this basis of building block objects to account for new structures that can appear at subleading power. 

The effective Lagrangian for hard scattering operators at any given order in the power counting, $ \cL^{(j)}_{\text{hard}}$, can be separated into a convolution between Wilson coefficients $\vec C$ encoding hard physics with $p^2\sim Q^2$, and on-shell physics encoded in SCET operators $\vec O$. In the hard scattering Lagrangian, the structure of SCET only allows convolutions between $\vec C$ and $\vec O$ in the collinear gauge invariant ${\cal O}(\lambda^0)$ momenta $\omega_i$, 
\begin{align} \label{eq:Leff_sub_explicit}
 \cL^{(j)}_{\text{hard}} = \sum_{\{n_i\}} \sum_{A,\{\lambda_j\}} 
 \bigg[ \prod_{i=1}^{\ell_{A}}\int\!\!\df \omega_i \bigg] \,
 & \vO^{(j)\dagger}_{A\{\lambda_j\}}\big(\{n_i\};
 \omega_1,\ldots,\omega_{\ell_A}\big) \,
 \vC^{(j)}_{A\{\lambda_j\}}\big(\{n_i\};\omega_1,\ldots,\omega_{\ell_A} \big)
 \,.
\end{align}
The operators $\vec{O}_A^{(j)}$ are traditionally constructed from the SCET building blocks in \eq{PC}, whereas here we will use helicity building blocks. The hard process being considered determines the appropriate collinear sectors $\{n_i\}$, and the relevant helicity combinations $\{\lambda_j\}$, which are a series of $\pm$s and $0$s, $\{\lambda_j\}=+-0+0+\cdots$. Different classes of operators are distinguished by the additional subscript $A$. which encodes all relevant information that is not distinguished by the helicity labels, such as particle content. This $A$ is also used to label the number of convolution variables $\ell_A$. The number of $\omega_i$'s depends on the specific operator we are considering since at subleading power multiple collinear fields can appear in the same collinear sector and we must consider the inclusion of ultrasoft building blocks with no $\omega_i$ labels. At leading power the operators $\vO^\dagger_{A\{\lambda_j\}}$ are given by products of the gluon and quark helicity building block operators in \eqs{cBpm_def}{jpm_def}.

The Wilson coefficients $\vC^{(j)}_{A\{\lambda_j\}}$ appearing in \Eq{eq:Leff_sub_explicit} are ${\cal O}(\lambda^0)$, and can be determined by a matching calculation. They are vectors in an appropriate color subspace. Since we will use building blocks that are simultaneously gauge invariant under collinear and ultrasoft transformations, the constraints of SCET gauge invariance are reduced to that of global color, making it simple to construct a color basis for these objects. Decomposing both the coefficients and operators in terms of color indices following the notation of~\cite{Moult:2015aoa}, we have
\begin{align} \label{eq:Cpm_Opm_color}
 C_{A\{\lambda_j\}}^{a_1\dotsb\alpha_n}
 & = \sum_k C_{A\{\lambda_j\}}^k T_k^{a_1\dotsb\alpha_n}
 \equiv \vT^{ a_1\dotsb\alpha_n} \vC_{A\{\lambda_j\}}
 \,, \nn\\
 \vO^\dagger_{A\{\lambda_j\}} &= \widetilde O_{A\{\lambda_j\}}^{a_1\dotsb \alpha_n}\, \vT^{\, a_1\dotsb \alpha_n}
 \,,
\end{align}
and the color space contraction in \eq{Leff_sub_explicit} becomes explicit, $\vec O_{A\{\lambda_j\}}^\dagger \, \vec C_{A\{\lambda_j\}} = \widetilde O_{A\{\lambda_j\}}^{a_1\dotsb \alpha_n}  C_{A\{\lambda_j\}}^{a_1\dotsb\alpha_n}$.
In \eq{Cpm_Opm_color} $\vT^{\, a_1\dotsb\alpha_n}$ is a row vector of color structures that spans the color conserving subspace. The $a_i$ are adjoint indices and the $\alpha_i$ are fundamental indices.  The color structures do not necessarily have to be independent, but must be complete. This issue is discussed in detail in~\cite{Moult:2015aoa}. Color structures which do not appear in the matching at a particular order will be generated by renormalization group evolution. (For a pedagogical review of the color decomposition of QCD amplitudes see~\cite{Dixon:1996wi,Dixon:2013uaa}.) 

In \SCETi, the leading power interactions between the soft and collinear degrees of freedom, described by $\cL^{(0)}$, can be decoupled using the BPS field redefinition \cite{Bauer:2002nz}
\be \label{eq:BPSfieldredefinition}
\cB^{a\mu}_{n\perp}\to \cY_n^{ab} \cB^{b\mu}_{n\perp} , \qquad \chi_n^\alpha \to Y_n^{\alpha \beta} \chi_n^\beta,
\ee
which is performed for fields in each collinear sector. Here $Y_n$, $\cY_n$ are fundamental and adjoint ultrasoft Wilson lines, respectively, and we note that
$Y_n T^a Y_n^\dagger =  T^b {\cal Y}_n^{ba}$.
For a general representation, r, the ultrasoft Wilson line is defined by
\be
Y^{(r)}_n(x)=\bold{P} \exp \bigg [ ig \int\limits_{-\infty}^0\!\!ds\: n\cdot A^a_{us}(x+sn)  T_{(r)}^{a}\bigg]\,,
\ee
where $\bold P$ denotes path ordering.  The BPS field redefinition generates ultrasoft interactions through the Wilson lines $Y_n^{(r)}$ which appear in the hard scattering operators~\cite{Bauer:2002nz}. When this is done consistently for S-matrix elements it accounts for the full physical path of ultrasoft Wilson lines~\cite{Chay:2004zn,Arnesen:2005nk}, so that some ultrasoft Wilson lines instead run over $[0,\infty)$. We can organize the result of this field redefinition by grouping the Wilson lines $Y_n^{(r)}$ together with elements in our color structure basis $\vT^{\, a_1\dotsb\alpha_n}$. We will denote the result of this by $\vT_{\rm BPS}^{\, a_1\dotsb\alpha_n}$.  As a simple leading power example of this, consider the operators
\be \label{eq:349}
O^{a\bar \alpha \beta}_{+(\pm)}= \cB_{1 +}^a\,J_{23\pm}^{\balpha\bt}\,,     \qquad     O^{a\bar \alpha \beta}_{-(\pm)}= \cB_{1 -}^a\,J_{23\pm}^{\balpha\bt} \,.
\ee
In this case there is a unique color structure before the BPS field redefinition, namely
\be \label{eq:gqqcolor}
\vT^{ a \al\bbeta} = \left ( T^a \right )_{\alpha \bbeta}\,.
\ee
After BPS field redefinition, we find the Wilson line structure,
\be \label{eq:example_BPS}
\vT_{\BPS}^{ a \al\bbeta} = Y^{\dagger \alpha \bar{\gamma}} _{n_2} T^b_{\gamma \bar{\sigma}} \cY_{n_1}^{ba}   Y_{n_3}^{\sigma \bbeta} \,.
\ee
The non-local structure encoded in these ultrasoft Wilson lines is entirely determined by the form of the operator in \Eq{eq:349}, and the definition of the BPS field redefinition in \Eq{eq:BPSfieldredefinition}.  After the BPS field redefinition, the building block fields are ultrasoft gauge invariant, but still carry global color indices. This will play an important role in defining gauge invariant helicity building blocks at subleading power, when ultrasoft fields appear in the hard scattering operators. In general we will use the notation
\begin{equation} \label{eq:Opm_BPScolor}
\vO^\dagger_{\{\lambda_j\}}
= O_{\{\lambda_j\}}^{a_1\dotsb \alpha_n}\, \vT_{\BPS}^{\, a_1\dotsb \alpha_n}
\,,
\end{equation}
for the operators with definite color indices that are obtained after the BPS field redefinition. After BPS field redefinition, $\vT_{\BPS}$ contains both color generators and ultrasoft Wilson lines, as in \Eq{eq:example_BPS}. This generalizes the vector of color structures used in the decomposition of the pre-BPS hard scattering operators in \Eq{eq:Cpm_Opm_color}, where to distinguish we included an extra tilde on the operators with specified color indices. More examples will be given in \Sec{sec:ang_cons}.

\section{Complete Set of Helicity Building Blocks}\label{sec:helops}

We now carry out the main goal of our paper, namely the extension of the scalar building blocks of \Eqs{eq:cBpm_def}{eq:jpm_def} to include all objects that are needed to describe subleading power interactions in the hard scattering Lagrangian. This will include defining operator building blocks involving multiple collinear fields in the same collinear sector, $\cP_\perp$ insertions, and explicit ultrasoft derivatives and fields.  We will continue to exploit the conservation of fermion number by organizing the fermions into bilinear currents. 

\begin{table}
 \begin{center}
  \begin{tabular}{|c|c|cc|ccc|c|ccc|}
	\hline \phantom{x} & \phantom{x} & \phantom{x} 
	& \phantom{x} & \phantom{x} & \phantom{x} & \phantom{x} 
	& \phantom{x} & \phantom{x} & \phantom{x} & \phantom{x} 
	\\[-13pt]                      
 Field: & 
    $\cB_{i\pm}^a$ & $J_{ij\pm}^{\balpha\beta}$ & $J_{ij0}^{\balpha\beta}$ 
    & $J_{i\pm}^{\balpha \beta}$ 
	& $J_{i0}^{\balpha \beta}$ & $J_{i\bar 0}^{\balpha \beta}$  
    & $\cP^{\perp}_{\pm}$ 
	& $\partial_{us(i)\pm}$ & $\partial_{us(i)0}$ & $\partial_{us(i)\bar{0}}$
	\\[3pt] 
 Power counting: &	
    $\lambda$ &  $\lambda^2$ &  $\lambda^2$
	& $\lambda^2$ & $\lambda^2$& $\lambda^2$ & $\lambda$ 
    & $\lambda^2$ & $\lambda^2$  & $\lambda^2$
	\\
 Equation: & 
   (\ref{eq:cBpm_def}) & \multicolumn{2}{c|}{(\ref{eq:jpm_def})} 
     & \multicolumn{3}{c|}{(\ref{eq:coll_subl})} & (\ref{eq:Pperppm}) 
     & \multicolumn{3}{c|}{(\ref{eq:partialus})}
    \\
  \hline  
  \end{tabular}\\
\vspace{.3cm} \hspace{2.85cm}
  \begin{tabular}{|c|cc|cccc|cc|}
	\hline  \phantom{x} &  \phantom{x} & \phantom{x} & \phantom{x} 
	 & \phantom{x} & \phantom{x} & \phantom{x} & \phantom{x} & \phantom{x}
	\\[-13pt]                        
 Field: & 
 	$\cB^a_{us(i)\pm}$ & 
    \!\!$\cB^a_{us(i)0}$ & 
	$J_{i(us)\pm}^{\balpha\beta}$  &
	$J_{i(\overline{us})\pm}^{\balpha\beta}$ &
	$J_{i(us)0}^{\balpha\beta}$ &
	$J_{i(\overline{us})0}^{\balpha\beta}$ &
	$J_{(us)^2ij\pm}$ & $J_{(us)^2ij0}$ 
	\\[3pt] 
 Power counting: &
 	$\lambda^2$ & $\lambda^2$ & $\lambda^4$ &  $\lambda^4$ &  $\lambda^4$
    & $\lambda^4$ & $\lambda^6$& $\lambda^6$
	\\ 
 Equation: & 
     \multicolumn{2}{|c|}{(\ref{eq:Bus})}  & \multicolumn{4}{c|}{(\ref{eq:Jus})} 
    & \multicolumn{2}{c|}{(\ref{eq:Jus2})} 
    \\
	\hline
  \end{tabular}
 \end{center}
\vspace{-0.3cm}
\caption{The complete set of helicity building blocks in $\text{SCET}_\text{I}$, together with their power counting order in the $\lambda$-expansion, and the equation numbers where their definitions may be found. The building blocks also include the conjugate currents $J^\dagger$ in cases where they are distinct from the ones shown.
} 
\label{tab:helicityBB}
\end{table}

A summary of our final results for the complete set of scalar building blocks valid to all orders in the \SCETi power expansion is shown in \Tab{tab:helicityBB}, along with the power counting of each building block and the equation number where it is defined. The building blocks that appeared already at leading power~\cite{Moult:2015aoa}, were given above in \eqs{cBpm_def}{jpm_def}. We will discuss each of the additional operators in turn.

For collinear gluons, the fields $\cB^a_{i\pm}$ suffice even at subleading power. An operator with an arbitrary number of collinear gluons in the same sector with arbitrary helicity and color indices can be formed by simply multiplying the $\cB^a_{i\pm}$ building blocks with the same collinear sector index $i$, such as $\cB^a_{i+}\cB^b_{i+}$.  On the other hand, for a quark-antiquark pair in the same collinear sector, the bilinear current building blocks of \eq{jpm_def} are not suitable. Indeed, the SCET projection relations
\begin{align} \label{eq:proj}
\frac{\Sl n_i \Sl {\bar n}_i}{4}  \chi_{n_i}=\chi_{n_i}, \qquad \Sl n_i \chi_{n_i}=0 
\,,
\end{align} 
enforce that the scalar current $\bar \chi_{n_i} \chi_{n_i}=0$, vanishes, as do the plus and minus helicity components of the vector current $\bar \chi_{n_i} \gamma_\perp^{\pm} \chi_{n_i}=0$.
In other words, the SCET projection relations enforce that a quark-antiquark pair in the same sector must have zero helicity if they are of the same chirality. Similarly, a quark-antiquark pair in the same sector with opposite chirality must have helicity $\pm 1$.
We therefore define the helicity currents 
\begin{align}\label{eq:coll_subl}
 & h=0:
 & J_{i0}^{\balpha \beta} 
  &= \frac{1}{2 \sqrt{\vphantom{2} \omega_{\bar \chi} \, \omega_\chi}}
  \: \bar \chi^\balpha_{i+}\, \Sl{\bar n}_i\, \chi^\beta_{i+}
   \,,\qquad
   J_{i\bar 0}^{\balpha \beta} 
  = \frac{1}{2 \sqrt{\vphantom{2} \omega_{\bar \chi} \, \omega_\chi}}
  \: \bar \chi^\balpha_{i-}\, \Sl {\bar n}_i\, \chi^\beta_{i-}
 \,, \\[5pt]
  & h=\pm 1:
 & J_{i\pm}^{\balpha \beta}
  &= \mp  \sqrt{\frac{2}{ \omega_{\bar \chi} \, \omega_\chi}}  \frac{\epsilon_{\mp}^{\mu}(n_i,\bar n_i)}{ \big(\l n_i \mp | \bar{n}_i \pm \r \big)^2}\: 
   \bar \chi_{i\pm}^\balpha\, \gamma_\mu \Sl{\bar n}_i\, \chi_{i\mp}^\beta
 \,. \nn
\end{align}
Because of the SCET projection relations of \Eq{eq:proj}, this set of currents, when combined with those of \Eq{eq:jpm_back_to_bacjdef} provides a complete set of building blocks for constructing hard scattering operators involving collinear fermions at all powers in the SCET expansion. Hard scattering operators involving arbitrary numbers of collinear quarks in different sectors, with arbitrary helicity and color indices, can be formed from products of these building blocks. The  $J_{i0}^{\balpha \beta}$ and $J_{i\bar 0}^{\balpha \beta}$ transform together as a scalar/pseudoscalar under rotations about the $\hat n_i$ axis, i.e. have helicity $h=0$. Similarly, the operators $J_{i\pm}^{\balpha \beta}$ have helicity $h=\pm 1$.  These four currents with quarks in the same collinear direction are shown in the second category in \Tab{tab:helicityBB}. These currents are again RPI-III invariant and our choice of prefactors is made to simplify their Feynman rules. The Feynman rules are simple to obtain, but we do not give them explicitly here. The Feynman rules for all currents in \SCETi and \SCETii will be given in~\cite{subhel:long}.  

Subleading power operators can also involve explicit insertions of the $\cP_{i\perp}^\mu$ operator.  Since the $\cP_{i\perp}^\mu$ operator acts on the perpendicular subspace defined by the vectors $n_i, \bar n_i$, which is spanned by the polarization vectors $\epsilon^{\pm}(n_i, \bar n_i)$, it naturally decomposes as 
\begin{align} \label{eq:Pperppm}
\cP_{i+}^{\perp}(n_i,\bar n_i)=-\epsilon^-(n_i,\bar n_i) \cdot \cP_{i\perp}\,, \qquad \cP_{i-}^{\perp}(n_i,\bar n_i)=-\epsilon^+(n_i,\bar n_i) \cdot \cP_{i\perp}\,.
\end{align} 
This decomposition is performed for the $\cP_{i\perp}$ operator in each sector. 
As we mentioned earlier, power counting ensures that the sector on which $\cP_{i\perp}$ acts is unambiguous. Hence we can simply drop the subscript $i$ and use $\cP^\perp_\pm$ as building blocks, as shown in \Tab{tab:helicityBB}. 

To see how this decomposition applies to operators written in more familiar notation, we consider the example operator $ \cP_\perp \cdot \cB_{i\perp}$. Using the completeness relation
\begin{align}\label{eq:completeness}
\sum\limits_{\lambda=\pm} \epsilon^\lambda_\mu(n_i,\bar n_i) \big[ \epsilon^\lambda_\nu(n_i,\bar n_i)  \big]^* = - g^\perp_{\mu \nu} ( n_i, \bn_i)\,,
\end{align}
the decomposition into our basis is given by
\begin{align}\label{eq:pdotbexample}
\cP_\perp \cdot \cB_{i\perp}=-\cP^{\perp}_{+} \cB_{i-}-\cP^{\perp}_{-} \cB_{i+}\,.
\end{align}
When acting within an operator containing multiple fields, square brackets are used to denote which fields are acted upon by the $\cP^{\perp}_{\pm}$ operator. For example
$\cB_{i+} \left [ \cP^{\perp}_{+}  \cB_{i-}  \right]  \cB_{i-}$,
indicates that the $\cP^{\perp}_{+}$ operator acts only on the middle field. Note that $\cP^\perp_\pm$ carry helicity $h=\pm 1$, and that the products in \eq{pdotbexample} behave like scalars.

To denote insertions of the $\cP^{\perp}_{\pm}$ operator into the currents of \Eq{eq:coll_subl} we establish a notation where the $\cP^{\perp}_{\pm}$ operator acts on only one of the two quark building block fields, by writing it either to the left or right of the current, and enclosing it in curly brackets. For example,
\begin{align}\label{eq:p_perp_notation}
  \big\{ \cP^{\perp}_\lambda J_{i 0 }^{\balpha \beta} \big\}  
  & = \frac{1}{2 \sqrt{\vphantom{2}\omega_{\bar \chi} \, \omega_\chi }} \:
   \Big[  \cP^{\perp}_{\lambda}  \bar \chi^\balpha_{i +}\Big] \Sl {\bar n}_i \chi^\beta_{i+}
  \,, \\
 \big\{ J_{i0 }^{\balpha \beta} (\cP^{\perp}_{\lambda})^\dagger \big\}
  &=  \frac{1}{2\sqrt{\vphantom{2}\omega_{\bar \chi} \, \omega_\chi}} \:
  \bar \chi^\balpha_{i+} \Sl {\bar n}_i \Big[   \chi^\beta_{i+} (\cP^{\perp}_{\lambda})^\dagger \Big]
  \,. \nn
\end{align}
If we wish to instead indicate a $\cP^\perp_\pm$ operator that acts on both building blocks in a current then we use the notation $\big[\cP^\perp_\lambda J_{i 0 }^{\balpha \beta}\big]$. The extension to multiple insertions of the $\cP^\perp_\pm$ operators should be clear.   Since the $\cP^\perp_\pm$ operators commute with ultrasoft Wilson lines, they do not modify the construction of the color bases either before or after the BPS field redefinition.

The operators defined in \Eq{eq:jpm_def}, \Eq{eq:coll_subl}, and \Eq{eq:Pperppm} form a complete basis of building blocks from which to construct hard scattering operators involving only collinear fields. As with the leading power operators, each of these subleading power operators is collinear gauge invariant, and therefore the treatment of color degrees of freedom proceeds as in \Eq{eq:Cpm_Opm_color}. Subleading hard scattering operators appearing in the ${\cal L}_{\rm hard}$ part of the SCET Lagrangian of \Eq{eq:SCETLagExpand} can be constructed simply by taking products of the scalar building blocks. Examples demonstrating the ease of this approach will be given in \Sec{sec:example}.

We now consider the remaining building blocks listed in \Tab{tab:helicityBB}, which all involve ultrasoft gluon fields, ultrasoft quark fields or the ultrasoft derivative operator $\partial_{us}$.  The simplicity of the collinear building blocks does not trivially extend to ultrasoft fields, since prior to the BPS field redefinition all collinear and ultrasoft objects transform under ultrasoft gauge transformations. This implies that constraints from ultrasoft gauge invariance must be imposed when forming an operator basis, and that the color organization of \Sec{sec:scethelicity} cannot be trivially applied to operators involving ultrasoft fields.  To overcome this issue, we can work with the hard scattering operators after performing the BPS field redefinition of \Eq{eq:BPSfieldredefinition}. The BPS field redefinition introduces ultrasoft Wilson lines, in different representations $r$, $Y^{(r)}_n(x)$, into the hard scattering operators. These Wilson lines can be arranged with the ultrasoft fields to define ultrasoft gauge invariant building blocks. The Wilson lines which remain after this procedure can be absorbed into the generalized color structure, $\vT_{\BPS}$, as was done at leading power in \Eq{eq:Opm_BPScolor}.

We begin by defining a gauge invariant ultrasoft quark field
\begin{align} \label{eq:usgaugeinvdef}
\psi_{us(i)}=Y^\dagger_{n_i} q_{us}\,,
\end{align}
where the direction of the Wilson line $n_i$ is a label for a collinear sector. Since the ultrasoft quarks themselves are not naturally associated with an external label direction, $n_i$ can be chosen arbitrarily, though there is often a convenient or obvious choice. This choice does not affect the result, but modifies the structure of the Wilson lines appearing in the hard scattering operators at intermediate stages of the calculation.  We also perform the following decomposition of the gauge covariant derivative in an arbitrary representation, $r$,
\begin{align}\label{eq:soft_gluon}
Y^{(r)\,\dagger}_{n_i} i D^{(r)\,\mu}_{us} Y^{(r)}_{n_i }=i \partial^\mu_{us} + [Y_{n_i}^{(r)\,\dagger} i D^{(r)\,\mu}_{us} Y^{(r)}_{n_i}]=i\partial^\mu_{us}+T_{(r)}^{a} g \cB^{a\mu}_{us(i)}\,,
\end{align}
where we have defined the ultrasoft gauge invariant gluon field by
\begin{align} \label{eq:softgluondef}
g \cB^{a\mu}_{us(i)}= \left [   \frac{1}{in_i\cdot \partial_{us}} n_{i\nu} i G_{us}^{b\nu \mu} \cY^{ba}_{n_i}  \right] \,.
\end{align}
In the above equations the derivatives act only within the square brackets. Again, the choice of collinear sector label $n_i$ here is arbitrary. This is the ultrasoft analogue of the gauge invariant collinear gluon field of \Eq{eq:chiBbare}, which can be written in the similar form
\begin{align}  
g\cB_{n_i\perp}^{A\mu} =\left [ \frac{1}{\bar \cP}    \bar n_{i\nu} i G_{n_i}^{B\nu \mu \perp} \cW^{BA}_{n_i}         \right]\,.
\end{align}
From the expression for the gauge invariant ultrasoft quark and gluon fields of \Eqs{eq:usgaugeinvdef}{eq:softgluondef} we see that unlike the ultrasoft fields, the operator $\cB^{A\mu}_{us(i)}$ is non-local at the scale $\lambda^2$, and depends on the choice of a collinear direction $n_i$. However the non-locality in our construction is entirely determined by the BPS field redefinition, and we can not simply insert arbitrary powers of dimensionless Wilson line products like $(Y_{n_1}^\dagger Y_{n_2})^k$ into the hard scattering operators. In practice this means that we can simply pick some $n_i$ for the Wilson lines in the building blocks in \eqs{usgaugeinvdef}{softgluondef} and then the BPS field redefinition determines the unique structure of remaining ultrasoft Wilson lines that are grouped with the color structure into  $\vT_{\BPS}^{a \al\bbeta}$. Determining a complete basis of color structures is straightforward. Detailed examples will be given in~\cite{subhel:long}, where the hard scattering operators for $e^+e^-\to$ dijets  involving ultrasoft fields will be constructed.

With the ultrasoft gauge invariant operators defined, we can now introduce ultrasoft fields and currents of definite helicity, which follow the structure of their collinear counterparts. Note from \eq{softgluondef}, that  $n_i\cdot \cB^{a}_{us(i)}= 0$.   For the ultrasoft gluon helicity fields we define the three building blocks
\begin{equation} \label{eq:Bus}
\cB^a_{us(i)\pm} = -\ve_{\mp\mu}(n_i, \bn_i)\,\cB^{a\mu}_{us(i)},\qquad  \cB^a_{us(i)0} =\bar n_\mu  \cB^{a \mu}_{us(i)}   
\,.\end{equation}
This differs from the situation for the collinear gluon building block in \eq{cBpm_def}, where only two building block fields were required, corresponding to the two physical helicities. For the ultrasoft gauge invariant gluon field we use three building block fields to describe the two physical degrees of freedom because the ultrasoft gluons are not fundamentally associated with any direction. Without making a further gauge choice, their polarization vectors do not lie in the perpendicular space of any fixed external reference vector.  If we use the ultrasoft gauge freedom to choose $\cB^a_{us(j)0}=0$, then we will still have $\cB^a_{us(i)0} \ne 0$ and $\cB^a_{us(i)\pm}\ne 0$ for $i\ne j$. We could instead remove $\cB^a_{us(j)0}$ for every $j$ using the ultrasoft gluon equation of motion, in a manner analogous to how $[W_{n_j}^\dagger i n_j\cdot D_{n_j} W_{n_j}]$ is removed for the collinear building blocks. However this would come at the expense of allowing inverse ultrasoft derivatives, $1/(in_j \cdot\partial_{us})$, to appear explicitly when building operators. While in the collinear case the analogous $1/\bnP$ factors are ${\cal O}(\lambda^0)$ and can be absorbed into the Wilson coefficients, this absorption would not be not possible for the ultrasoft case. Therefore, for our \SCETi construction we choose to forbid explicit inverse ultrasoft derivatives that can not be moved into Wilson lines, and allow $\cB^a_{us(i)0}$ to appear. An example of a case where the non-locality can be absorbed is given in \eq{softgluondef}, where the $1/(in\cdot\partial_{us})$ is absorbed into ultrasoft Wilson lines according to \eq{soft_gluon}. Thus the only ultrasoft non-locality that appears in the basis is connected to the BPS field redefinition.

We also decompose the ultrasoft partial derivative operator $\partial_{us}^\mu$ into lightcone components,
\begin{equation}  \label{eq:partialus}
\partial_{us(i)\pm} = -\ve_{\mp\mu}(n_i, \bn_i)\,\partial^{\mu}_{us},\qquad   \partial_{us(i)0} =\bar n_{i\mu} \partial^{\mu}_{us}, \qquad \partial_{us(i)\bar 0} = n_{i \mu} \partial^{\mu}_{us}
\,.\end{equation}
In contrast with the collinear case, we cannot always eliminate the $n_i \cdot \partial_{us}$ using the equations of motion without introducing inverse ultrasoft derivatives (e.g. $1/(\bn_i \cdot \partial_{us})$) that are unconnected to ultrasoft Wilson lines. When inserting ultrasoft derivatives into operators we will use the same curly bracket notation defined for the $\cP_\perp$ operators in \Eq{eq:p_perp_notation}. In other words, $\{i \partial_{us(i)\lambda} J\}$ indicates that the ultrasoft derivative acts from the left on the first field in $J$ and $\{  J (i\partial_{us(i)\lambda})^\dagger\}$ indicates that it acts from the right on the second field in $J$.

Gauge invariant ultrasoft quark fields also appear explicitly in the operator basis at subleading powers. Due to fermion number conservation they are conveniently organized into scalar currents. From \eq{PC}, we see that ultrasoft quark fields power count like $\lambda^3$. However, for factorization theorems involving a single collinear sector, as arise when describing a variety of inclusive and exclusive $B$ decays (see e.g.  ~\cite{Manohar:1993qn,Beneke:2000ry,Beneke:2000wa,Bauer:2000yr,Beneke:2001at,Bauer:2001ct,Bauer:2001cu,Bosch:2001gv,Bauer:2001yt,Bauer:2002aj,Beneke:2003pa,Beneke:2004dp,Arnesen:2005ez,Lee:2005pwa,Lee:2006gs}), operators involving ultrasoft quarks appear at leading power. The currents involving both collinear and ultrasoft quarks that are necessary to define subleading power operators at any desired order are 
\begin{align} \label{eq:Jus}
 J_{i(us)\pm}^{\balpha\beta}
  &= \mp  \:
  \frac{\ve_\mp^\mu(n_i, \bar n_i)}{ \l \bn_i \mp | n_i \pm \r}\: 
  \bar{\chi}^\balpha_{i\pm}\,  \gamma_\mu \psi^\beta_{us(i)\pm}
 \,,  \\
 J_{i(\overline{us})\pm}^{\balpha\beta}
  &=\mp  
  \frac{\ve_\mp^\mu(\bar{n}_i, n_i)}{ \l n_i \mp | \bar{n}_i \pm \r}\: \bar{\psi}_{us(i) \pm}^\balpha\, \gamma_\mu \chi^\beta_{i\pm} 
 \, , \nn \\
 J_{i(us)0}^{\balpha\beta}
  &=   \bar \chi^\balpha_{i+}\psi^\beta_{us(i)-}
  \,, \qquad\qquad\qquad
  (J^\dagger)_{i(us)0}^{\balpha\beta}
  =   \bar \psi^\balpha_{us(i)-} \chi^\beta_{i+}
  \,, \nn\\
 J_{i(\overline{us})0}^{\balpha\beta}
  &=  \bar\psi^\balpha_{us(i)+} \chi^\beta_{i-}
  \,, \qquad\qquad\qquad
  (J^\dagger)_{i(\overline{us})0}^{\balpha\beta}
  = \bar \chi^\balpha_{i-}\psi^\beta_{us(i)+}
  \,, \nn
\end{align}
For these mixed collinear-ultrasoft currents we choose to use the collinear sector label $i$ in order to specify the ultrasoft quark building block field. In addition, we need currents that are purely built from ultrasoft fields,
\begin{align}  \label{eq:Jus2}
 J_{(us)^2 ij\pm}^{\balpha\beta}
  &= \mp\, \frac{\ve_\mp^\mu(n_i, n_j)}{\langle n_j\mp | n_i\pm\rangle}\: 
  \bar{\psi}^\balpha_{us(i)\pm} \gamma_\mu \psi^\beta_{us(j)\pm} 
  \,, \\
 J_{(us)^2 ij 0}^{\balpha\beta}
  &=
  \bar \psi^\balpha_{us(i)+}\psi^\beta_{us(j)-}
  \,,\qquad\qquad\quad
 (J^\dagger)_{(us)^2 ij 0}^{\balpha\beta}
  =
  \bar \psi^\balpha_{us(i)-}\psi^\beta_{us(j)+}
  \,. \nn
\end{align}
To specify the building blocks in these ultrasoft-ultrasoft currents we use two generic choices, $i$ and $j$, with $n_i \neq n_j$ so as to make the polarization vector well defined. Although the ultrasoft quark carries these labels, they are only associated with the Wilson line structure and, for example, the ultrasoft quark building block fields do not satisfy the projection relations of \eq{proj}.  

The ultrasoft currents in \eq{Jus} complete our construction of the complete set of scalar building blocks given in \Tab{tab:helicityBB}.  The objects in this table can be used to construct bases of hard scattering operators at any order in the power counting parameter $\lambda$, by simply taking products of the scalar building blocks.


There are several extensions to this construction that should be considered. One is the extension to \SCETii with collinear and soft fields, rather than collinear and ultrasoft fields. A table of scalar building block operators for \SCETii that is analogous to \Tab{tab:helicityBB} will be given in~\cite{subhel:long}.  Also, the completeness of the set of helicity building blocks relies on massless quarks and gluons having two helicities, which is specific to $d=4$ dimensions. Depending on the regularization scheme, this may or may not be true when dimensional regularization with $d=4-2\epsilon$ dimensions is used, and evanescent operators \cite{Buras:1989xd,Dugan:1990df,Herrlich:1994kh}, beyond those given in \Tab{tab:helicityBB} can appear. While evanescent operators are not required at leading power, (see~\cite{Moult:2015aoa} for a detailed discussion), this need no longer be the case at subleading power, and will be discussed further in~\cite{subhel:long}.

\section{Constraints from Angular Momentum Conservation}\label{sec:ang_cons}

If we include the spin of objects that are not strongly interacting, such as electrons and photons, then the overall hard scattering operators in \eq{Leff_sub_explicit} are scalars under the Lorentz group. In this section we will show that this constraint on the total angular momentum gives restrictions on the angular momentum that is allowed in individual collinear sectors. These restrictions become nontrivial beyond leading power, when multiple operators appear in the same collinear sector.

If we consider a leading power hard scattering process where two gluons collide to produce two well separated quark jets plus an $e^+e^-$ pair, then this is described by a leading power operator with each field sitting alone in a well separated collinear direction, such as
\begin{align} \label{eq:BBJJee}
   \cB_{1\lambda_1}^a \cB_{2\lambda_2}^b J_{34\lambda_q}^{\balpha\beta} J_{e56\lambda_e} \,.
\end{align}
Here, the leading power electron current is defined in a similar way as the quark current, but without gluon Wilson lines,
\begin{equation}
J_{e \pm } \equiv 
J_{e ij\pm }
= \mp \sqrt{\frac{2}{\omega_i\, \omega_j}}\, \ve_\mp^\mu(n_i, n_j)\,    \frac{\bar{e}_{i\pm} \gamma_\mu e_{j \pm}} {\langle n_j\mp | n_i\pm\rangle}
\,.
\end{equation}
For notational convenience we will drop the explicit $ij$ label on the electron current, denoting it simply by $J_{e\pm}$. Although the operator in \eq{BBJJee} has to be a scalar, there are still no constraints on the individual values of the $\lambda_i$. Each building block has spin components that are defined with respect to a distinct axis $\hat n_i$, and yields a linear combination of spin components when projected onto a different axis. Thus, projecting all helicities onto a common axis we only find the trivial constraint that the angular momenta factors of $1$ or $1/2$ from each sector must together add to zero.\footnote{There are of course simple examples where this constraint reduces the basis of operators. For example, for gluon fusion Higgs production, angular momentum conservation implies that only two operators are required in the basis
\begin{align}
O_{++}^{ab}
= \frac{1}{2}\, \cB_{1+}^a\, \cB_{2+}^b\,  H_3
\,, \qquad
O_{--}^{ab}
= \frac{1}{2}\, \cB_{1-}^a\, \cB_{2-}^b\, H_3\nn
\,,\end{align}
where $H_3$ is the scalar Higgs field.
}
In the example of \eq{BBJJee}, this is $1\oplus 1\oplus\frac12\oplus\frac12\oplus\frac12\oplus\frac12 =0$ for a generic kinematic configuration.\footnote{If we were in a frame where the gluons were back-to-back, there spins would be combined along a single axis. In this example, this would still not give us any additional restrictions.} Note that for the quark and electron currents here, we have individual spin-$1/2$ fermions in different directions, so $\lambda_q$ and $\lambda_e$ do not correspond to helicities. As another example, consider $4$-gluon scattering, with all gluon momenta well separated and thus in their own collinear sectors, we have the operators 
\begin{align}
\cB_{1\lambda_1}^{a_1}\cB_{2\lambda_2}^{a_2}\cB_{3\lambda_3}^{a_3}\cB_{4\lambda_4}^{a_4}\,. 
\end{align}
Here we can again specify the helicities $\lambda_i=\pm$ independently, because each of these helicities is specified about a different quantization axis.  Each carries helicity $h=\pm1$, and angular momentum is conserved because these four spin-$1$ objects can add to spin-$0$.  Therefore all helicity combinations must be included.

To understand the constraints imposed by angular momentum conservation at subleading power, it is interesting to consider a specific example in more detail. As a simple example, consider an $e^+e^-$ collision in the center of mass frame producing two back-to-back jets, where we label the associated jet directions as $n$ and $\bar n$. The leading power operators are
\begin{align}\label{eq:LP_basis_angsec}
O_{(+;+)}^{(0)\balpha\bt}
=J^{\balpha\bt}_{n \bar n+}J_{e +}\,, \qquad
O_{(+;-)}^{(0)\balpha\bt}
=J^{\balpha\bt}_{n \bar n+}J_{e -}\,,  \\
O_{(-;+)}^{(0)\balpha\bt}
=J^{\balpha\bt}_{n \bar n-}J_{e +}\,, \qquad
O_{(-;-)}^{(0)\balpha\bt}
=J^{\balpha\bt}_{n \bar n-}J_{e -}\,,\nn
\end{align}
where $J_{n\bn\pm}^{\balpha\bt}$ were defined in \eq{jpm_back_to_bacjdef}.
Here, we can view $J^{\balpha\bt}_{n \bar n \pm}$ as creating or destroying a state of helicity $h=\pm1$ about the $n$ axis, and $J_{e\pm}$ as creating or destroying a state of helicity $h=\pm 1$ about the electron beam axis. Defining $\theta$ as the angle between the quark and electron and taking all of the particles to be outgoing, the spin projection implies that the Wilson coefficients are proportional to the Wigner $d$ functions,
\begin{alignat}{2}
&C_{(+;+)}^{(0)\balpha\bt}
\propto 1+\cos \theta\,, \qquad & &
C_{(+;-)}^{(0)\balpha\bt}
\propto 1-\cos \theta   \,,  \\
&C_{(-;+)}^{(0)\balpha\bt}
\propto 1-\cos \theta  \,, \qquad &&
C_{(-;-)}^{(0)\balpha\bt}
\propto 1+\cos \theta \,.\nn
\end{alignat}
As expected, all helicity combinations are non-vanishing (except when evaluated at special kinematic configurations).

\begin{figure}[t!]
%
%
\begin{center}
\includegraphics[width=0.35\columnwidth]{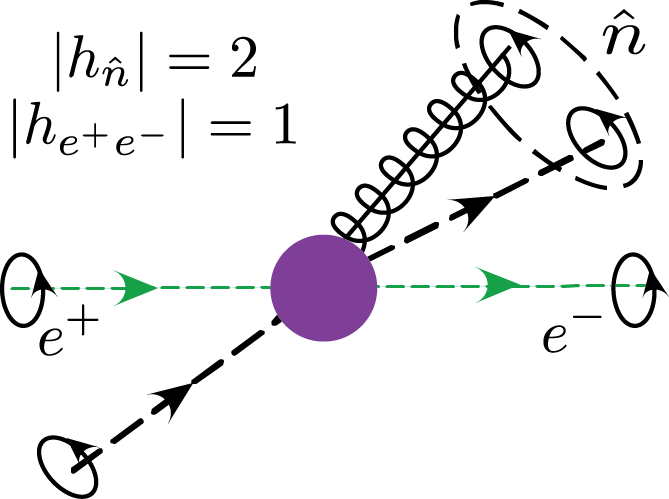} 
\hspace{1.4cm}
\includegraphics[width=0.35\columnwidth]{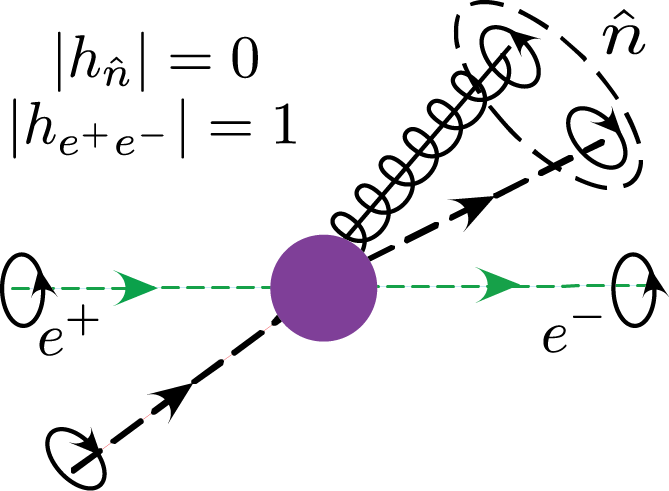} 
\raisebox{0cm}{ \hspace{-3.0cm} 
  $a$)\hspace{6.6cm}
  $b$)\hspace{2.4cm} }
\\[-25pt]
\end{center}
\vspace{-0.4cm}
\caption{ 
A schematic illustration of the helicity selection rule with two axes, as relevant for the case of $e^+e^-\to $ dijets. In a) the $n$-collinear sector carries $|h|=2$, and therefore has a vanishing projection onto the $J_{e\pm}$ current. In b), the collinear sector carries $|h|=0$ and has a non-vanishing projection onto the $J_{e\pm}$ current. } 
\label{fig:helicity_constraint}
\end{figure}

Considering this same example at subleading power, the analysis of angular momentum becomes more interesting, since multiple fields are present in a single collinear sector. For the subleading $e^+e^-\to $ dijet operators with only $n$-collinear and $\bn$-collinear fields, we only have a single axis $\hat n$ for all strongly interacting operators, and can simply add up their helicities to determine the helicity $h_{\hat n}$ in this direction.  Since the operator in the only other direction, $J_{e\pm }$, has spin-1, this implies that the total helicity for the $n$-$\bar n$ sector must be $h_{\hat n}=0,1,-1$ for the operator to have a non-vanishing contribution. Any operator with $|h_{\hat n}|>1$ must belong to a representation of spin $J>1$, and is ruled out because we can not form a scalar when combining it with the spin-$1$ electron current. An example of this is shown in \Fig{fig:helicity_constraint}.

As an explicit example of the constraints that this places on the subleading power helicity operators, consider the $\mathcal{O}(\lambda)$ back-to-back collinear operators  involving two collinear quark fields and a single collinear gluon field, which appears at $\cO(\lambda)$. For the case that the quarks are in different collinear sectors we can start by considering the operator list
\begin{alignat}{2} \label{eq:Z1_basis_cons}
&O_{+(+;\pm)}^{(1)a\,\balpha\bt}
=\cB_{n+}^a \, J_{n\bar n\,+}^{\balpha\bt}\,  J_{e\pm }
\,,\qquad &
&O_{+(-;\pm)}^{(1)a\,\balpha\bt}
= \cB_{n+}^a\, J_{n\bar n\,-}^{\balpha\bt}\, J_{e\pm }
\,,  \\
&O_{-(+;\pm)}^{(1)a\,\balpha\bt}
= \cB_{n-}^a\, J_{n\bar n\,+}^{\balpha\bt}\, J_{e\pm }
\,,\qquad &
&O_{-(-;\pm)}^{(1)a\,\balpha\bt}
= \cB_{n-}^a \, J_{n\bar n\, -}^{\balpha\bt}\,J_{e\pm }
\,, \nn
\end{alignat}
while for the case that the quarks are in the same collinear sector we consider
\begin{alignat}{2} \label{eq:Z1_basis_diff_cons}
 &O_{\bar{n}+(0;\pm)}^{(1)a\,\balpha\bt}
= \cB_{n+}^a\, J_{\bar n0\, }^{\balpha\bt}\,J_{e\pm }
\,,\qquad &
&O_{\bar{n}+(\bar 0;\pm)}^{(1)a\,\balpha\bt}
=  \cB_{n+}^a\, J_{\bar n\bar 0\, }^{\balpha\bt}\,J_{e\pm }
\,,\\
&O_{\bar{n}-(0;\pm)}^{(1)a\,\balpha\bt}
= \cB_{n-}^a\,  J_{\bar n0\, }^{\balpha\bt}\,J_{e\pm }
\,,\qquad &
&O_{\bar{n}-(\bar 0;\pm)}^{(1)a\,\balpha\bt}
=\cB_{n-}^a\, J_{\bar n\bar 0\, }^{\balpha\bt} \,J_{e\pm }
\,.\nn
\end{alignat}
We have used the fact that chirality is conserved in massless QCD, eliminating the need to consider $J_{n\bn 0}^{\balpha\bt}$ or $J_{\bn \pm}^{\balpha\bt}$ for the process being considered here. There are also operators with $\cB_{\bn\pm}^a$ that are obtained from those in \eqs{Z1_basis_cons}{Z1_basis_diff_cons} by taking $n\leftrightarrow \bn$. Furthermore, we do not consider the color structure, as it is irrelevant for the current discussion.  (Also note that we are not attempting to enumerate all ${\cal O}(\lambda)$ operators here. This is done in~\cite{subhel:long}.)

The constraint from conservation of angular momentum gives further restrictions, implying that only a subset of the eight operators in \eqs{Z1_basis_cons}{Z1_basis_diff_cons} are non-vanishing. In \eq{Z1_basis_cons} the strongly interacting operators have $h_{\hat n}=0$ or $h_{\hat n}=\pm2$,  and only those with $h_{\hat n}=0$ can contribute to the $J=0$ hard scattering Lagrangian, leaving only
\begin{alignat}{2} \label{eq:Z1_basis_cons_actuallyconserved}
&O_{+(-;\pm)}^{(1)a\,\balpha\bt}
= \cB_{n+}^a\, J_{n\bar n\,-}^{\balpha\bt}\,  J_{e\pm }
\,,\qquad &
&O_{-(+;\pm)}^{(1)a\,\balpha\bt}
= \cB_{n-}^a\, J_{n\bar n\,+}^{\balpha\bt}\,J_{e\pm} \,.
\end{alignat}
Thus angular momentum reduces the number of hard scattering operators by a factor of two in this case. On the other hand, for the case with both quarks in the same collinear sector in \eq{Z1_basis_diff_cons}, the operators all have $h_{\hat n}=\pm1$, and therefore all of them are allowed. 

Having understood how the angular momentum conservation constraint appears in the helicity operator language, it is interesting to examine how it appears if we instead work with the traditional operators of \eq{PC}. 
Here we must construct the SCET currents $\mathcal{J}^\mu$ at $\mathcal{O}(\lambda)$ involving two collinear quarks and a collinear gluon. The Lorentz index on $\mathcal{J}^\mu$ is contracted with the leptonic tensor to give an overall scalar, and thus preserve angular momentum. The operators in a basis for $\mathcal{J}^\mu$ can be formed from Lorentz and Dirac structures, as well as the external vectors, $n^\mu$ and $\bar n^\mu$. When the collinear quarks are each in a distinct collinear sector, the SCET projection relations of \Eq{eq:proj} imply that $\bar \chi_{\bar n} {\Sl {n}} \chi_n =  \bar \chi_{\bar n} {\Sl {\bar n}} \chi_n =0$. To conserve chirality we must have a $\gamma^\perp_\nu$ between the quark building blocks, and this index must be contracted with the other free $\perp$-index, $\nu$, in the collinear gluon building block $\cB_{\perp n}^\nu$ (which we again choose to be in the $n$ direction). Therefore an $n$ or $\bar n$ must carry the $\mu$ Lorentz index. After the BPS field redefinition it can be shown\footnote{Note that in constructing a complete basis of Lorentz and Dirac structures for \eqs{dirac_op_1}{dirac_op_2}, that all other operators can be eliminated using symmetry properties and the conservation of the current,  $q_\mu \mathcal{J}^{(1)\mu}_i=0$. Eliminating operators here is tedious compared to the helicity operator approach.} that for photon exchange the unique $\mathcal{O}(\lambda)$ operator with collinear quark fields in distinct collinear sectors is 
\begin{align}\label{eq:dirac_op_1}
\mathcal{J}^{(1)\mu}_1 =r^\mu_-\bar \chi_{\bar n}Y^\dagger_{\bar n}Y_n \Sl\cB_{\perp n} \chi_{n}\,,
\end{align}
where, defining $q^\mu$ as the sum of the momenta of the colliding leptons, we have
\begin{align}
r^\mu_-= \frac{n \cdot q}{2} \bar n^\mu -\frac{\bar n \cdot q}{2} n^\mu\,.
\end{align}
In the case that both collinear quark fields are in the same collinear sector, similar arguments using the SCET projection relations can be used to show that the collinear gluon field must carry the Lorentz index, and that the unique operator is 
\begin{align} \label{eq:dirac_op_2}
\mathcal{J}^{(1)\mu}_2 =\bar \chi_{\bar n}Y^\dagger_{\bar n}Y_n\cB_{ \perp n}^{\mu}Y^\dagger_{n} Y_{\bar n}\Sl r_{- } \chi_{\bar n} \,.
\end{align}

We see a direct correspondence between \Eqs{eq:Z1_basis_diff_cons}{eq:dirac_op_2}. In both equations the collinear quark fields have $h=0$ and thus form a scalar, and the collinear gluon field carries the spin that is combined with the leptonic current. For photon exchange, all of the Wilson coefficients of the operators in \eq{Z1_basis_diff_cons} are related by CP properties and angular momentum constraints, so there is only one combination of the four operators that appears with a nontrivial Wilson coefficient. This combination maps exactly to the single operator in \eq{dirac_op_2}.  We also see a correspondence between \Eqs{eq:Z1_basis_cons_actuallyconserved}{eq:dirac_op_1}, where both collinear quarks are contracted with the collinear gluon to form a $h=0$ combination. Indeed, using the completeness relation of \Eq{eq:completeness} for $g^\perp_{\mu \nu} ( n_i, \bn_i)$, the operators of \Eqs{eq:dirac_op_1}{eq:dirac_op_2} can straightforwardly be converted to the helicity operators of \Eqs{eq:Z1_basis_diff_cons}{eq:Z1_basis_cons_actuallyconserved}. 

It is interesting to note that when working in terms of building blocks involving Lorentz and Dirac structures, the SCET projection relations, which were ultimately what allowed us to define helicity fields along given axes, played a central role in reducing the basis. One is also forced to incorporate the constraints from the total angular momentum as part of the analysis, by the need to keep track of the contraction of Lorentz indices. In the helicity operator basis the same constraints appear as simple elimination rules on the allowed helicities when taking products of building blocks in the same collinear sector (and any back-to-back sector if one is present). These products can be classified by the minimal total angular momentum object for which they are a component, and eliminated if this value is too large.

We can now specify the general constraint from angular momentum on the helicities of an operator basis. The operator basis must be formed such that $J^{(i)}_{\rm min}$, the minimal angular momentum carried by the $n_i$-collinear sector,  satisfies 
\begin{align} \label{eq:Jmin}
J^{(i)}_\text{min}\leq \sum_{j \text{ with }  \hat n_j\neq \hat n_i} J_{\rm min}^{(j)} \,.
\end{align}
If the helicities in the $n_i$-collinear sector of some operator add up to $h_{n_i}^{\rm tot}$, then the minimum angular momentum for that sector is $J^{(i)}_\text{min}=|h_{n_i}^{\rm tot}|$. Therefore we can write \eq{Jmin} in a form that is useful for constraining the helicity of operators,
\begin{align} \label{eq:hmin}
 |h_{n_i}^\text{tot}| \leq \sum_{j \text{ with }  \hat n_j\neq \hat n_i}  | h_{n_j}^\text{tot}|  \,.
\end{align}
In cases where two of our light-like vectors are back-to-back, $n_i\cdot n_{k} =2 +{\cal O}(\lambda^2)$, then the operators in both the $n_i$ and $n_k$ collinear directions are considered simultaneously when calculating the value of $h_{n_i}^\text{tot}$ (where $\pm$ for $n_k$ count as $\mp$ for $n_i$), and not as distinct terms in the sum. This includes the case where $n_k=\bn_i$.  \eq{hmin} prevents subleading power operators from having exceedingly large angular momenta about any particular collinear direction.

This constraint of angular momentum conservation of the hard scattering process shows that when writing down a basis of helicity operators, not all helicity combinations should be included in the basis. Especially when working at higher powers, this places considerable constraints on the basis, and supplements additional constraints from parity and charge conjugation invariance (see~\cite{Moult:2015aoa}). This reduction can be contrasted with the leading power operators explored in~\cite{Moult:2015aoa}, where most often all possible different helicity combinations had to be included in the basis of hard scattering operators.

\section{Example: $q\bar qgg$ Operators for $n$-$\bn$ Directions}\label{sec:example}

To demonstrate the simplicity of the helicity operator approach, in this section we  will explicitly construct a basis of hard scattering operators with two back-to-back collinear sectors, $n$ and $\bn$. For simplicity, we will restrict ourselves to the channel involving two collinear gluons, a collinear quark and a collinear antiquark.  The operators to be discussed in this section are suppressed by $\cO(\lambda^2)$ compared to the leading power operator, which involves a quark and antiquark field in opposite collinear sectors, and contribute at subleading power to $e^+ e^- \to$ dijet event shapes, Drell-Yan, or DIS with one jet. They do not in themselves constitute a complete basis of ${\cal O}(\lambda^2)$ operators, but do make up a unique subset which we can use to illustrate the power of our approach. The complete ${\cal O}(\lambda^2)$ basis of operators will be presented and analyzed in~\cite{subhel:long}.

The angular momentum arguments of \sec{ang_cons} enforce that the helicity along the single jet axis satisfies $|h_{\hat n}^{\text{tot}}| \le 1$. Additionally, for the particular process $e^+e^-\to$ dijets the quark and antiquark have the same chirality, which provides further restrictions on the allowed operators that we will enumerate below. Using the notation of \eq{Cpm_Opm_color} we write the three-dimensional color basis for the $q\bar q gg$ channels as
\begin{equation} \label{eq:ggqqll_color}
\vT^{\, ab \alpha\bbeta}
= \Bigl(
(T^a T^b)_{\alpha\bbeta}\,,\, (T^b T^a)_{\alpha\bbeta} \,,\, \tr[T^a T^b]\, \delta_{\alpha\bbeta}
\Bigr)
\,.\end{equation} 
The color basis after BPS field redefinition will be given separately for each distinct partonic configuration, each of which will be discussed in turn.

We begin by considering operators where the quark and antiquark fields have distinct collinear sector labels, and the gluon fields are in the same collinear sector. In this case, a basis of helicity operators is
\begin{align}
 \boldsymbol{(gg q)_n (\bar q)_{\bn}:}     {\vcenter{\includegraphics[width=0.18\columnwidth]{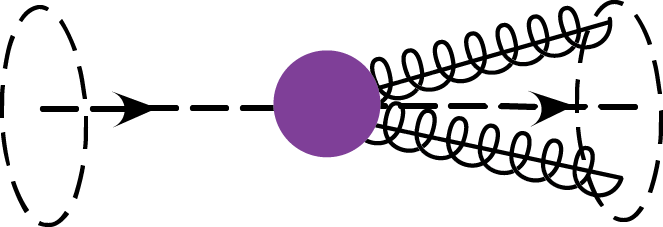}}}
\nn
\end{align}
\vspace{-0.4cm}
\begin{alignat}{2} \label{eq:eeqqgg_basis1}
&O_{\cB1++(-;\pm)}^{(2)ab\, \balpha\bt}
= \frac{1}{2}   \cB_{n+}^a\, \cB_{n+}^b\, J_{n\bar n\, - }^{\balpha\bt}   J_{e\pm }
\, , \qquad 
&&O_{\cB1--(+;\pm)}^{(2)ab\, \balpha\bt}
= \frac{1}{2}  \cB_{ n-}^a\, \cB_{ n-}^b \, J_{n\bar n\, + }^{\balpha\bt}   J_{e\pm }
\, ,  \nn\\
&O_{\cB1+-(+;\pm)}^{(2)ab\, \balpha\bt}
=  \cB_{n+}^a\, \cB_{n-}^b  \, J_{n\bar n\, +}^{\balpha\bt}   J_{e\pm }
\, , \qquad 
&&O_{\cB1+-(-;\pm)}^{(2)ab\, \balpha\bt}
=  \, \cB_{n+}^a\, \cB_{ n-}^b J_{n\bar n\, -}^{\balpha\bt}    J_{e\pm }
\, , 
\end{alignat}
\vspace{2 cm}
\begin{align}
 \boldsymbol{(gg \bar q)_n (q)_{\bn}:}   {\vcenter{\includegraphics[width=0.18\columnwidth]{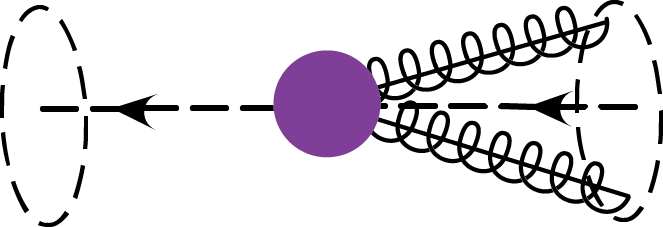}}}
\nn
\end{align} 
\vspace{-0.4cm} 
\begin{alignat}{2} \label{eq:eeqqgg_basis1a}
&O_{\cB2++(-;\pm)}^{(2)ab\, \balpha\bt}
= \frac{1}{2}   \cB_{n+}^a\, \cB_{n+}^b\, J_{\bar n n\, + }^{\balpha\bt}   J_{e\pm }
\, , \qquad 
&&O_{\cB2--(+;\pm)}^{(2)ab\, \balpha\bt}
= \frac{1}{2}  \cB_{ n-}^a\, \cB_{ n-}^b \, J_{\bar n n\, - }^{\balpha\bt}   J_{e\pm }
\, , \nn \\
&O_{\cB2+-(+;\pm)}^{(2)ab\, \balpha\bt}
=  \cB_{n+}^a\, \cB_{n-}^b  \, J_{\bar n n\, +}^{\balpha\bt}   J_{e\pm }
\, , \qquad 
&&O_{\cB2+-(-;\pm)}^{(2)ab\, \balpha\bt}
=  \, \cB_{n+}^a\, \cB_{ n-}^b J_{\bar n n\, -}^{\balpha\bt}    J_{e\pm }
\, . 
\end{alignat}
Here we have used constraints from angular momentum conservation to eliminate operators whose non-leptonic component do not have $h=0,\pm1$ along the $\hat n$ axis. For example, we have not allowed the operators $ \cB_{n+}^a\, \cB_{n+}^b\, J_{n\bar n\, + }^{\balpha\bt}   J_{e\pm }$ which have $h=+3$ along the $n$ axis and could not be created from the intermediate vector boson. Also, we have used the $n \leftrightarrow \bn$ symmetry to only write operators with both gluons in the $n$-collinear sector, a simplification that we will make repeatedly in this section. Operators with $\bn$-collinear gluons are obtained by simply taking $n \leftrightarrow \bn$. The color basis for the operators in \eqs{eeqqgg_basis1}{eeqqgg_basis1a} after the BPS field redefinition is
\begin{equation} \label{eq:BPS1}
\vT_{\BPS}^{\, ab \alpha\bbeta}
= \Bigl(
(T^a T^bY^\dagger_n Y_{\bar n})_{\alpha\bbeta}\,,\, (T^b T^aY^\dagger_n Y_{\bar n})_{\alpha\bbeta} \,,\, \tr[T^a T^b]\, [Y^\dagger_n Y_{\bar n}]_{\alpha\bbeta}
\Bigr)
\,.
\end{equation}
In order to see how this is derived, we will go through the algebra explicitly for the first color structure. Using the result for the transformations in \eq{BPSfieldredefinition}, we see that each gluon field from (\ref{eq:eeqqgg_basis1}) or (\ref{eq:eeqqgg_basis1a}) contributes an adjoint Wilson line while each fermion contributes a fundamental Wilson line. So, our color structure becomes
\begin{align} \label{eq:color-BPS-explicit}
(T^aT^b)_{\alpha \bbeta} &\to (Y_n^\dagger T^{a'} \cY_n^{a' a} T^{b'} \cY_n^{b' b} Y_{\bn})_{\alpha \bbeta} = (Y_n^\dagger Y_n T^{a} Y_n^{\dagger} Y_n T^{b} Y_n^{\dagger} Y_{\bn})_{\alpha \bbeta} \nn \\
&= (T^a T^bY^\dagger_n Y_{\bar n})_{\alpha\bbeta}\,,
\end{align}
where we have used $T^{a'} \cY_i^{a' a} = Y_i T^a Y^\dagger_i$. Similar manipulations give the other Wilson line structures in \eq{BPS1}.

Next we consider the operators where the quark and antiquark fields have distinct collinear sector labels, as do the gluons. In this case, the basis of helicity operators is
\begin{align}
 \boldsymbol{(g q)_n (g \bar q)_{\bn}:}   {\vcenter{\includegraphics[width=0.18\columnwidth]{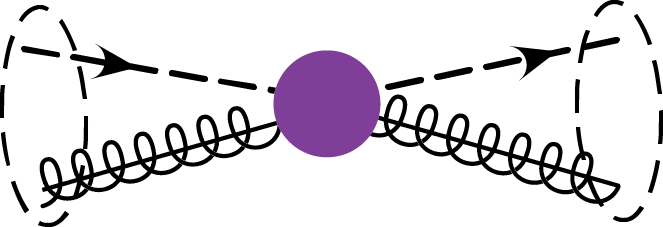}}}  \nn
\end{align}
\vspace{-0.45cm}
\begin{alignat}{2} \label{eq:eeqqgg_basis2}
&O_{\cB3++(+;\pm)}^{(2)ab\, \balpha\bt}
= \cB_{n+}^a\, \cB_{\bar n+}^b \, J_{n\bar n\, + }^{\balpha\bt}   J_{e\pm } \, , \qquad 
&&O_{\cB3--(-;\pm)}^{(2)ab\, \balpha\bt}
=    \cB_{n-}^a\, \cB_{\bar n-}^b \, J_{n\bar n\, -}^{\balpha\bt} J_{e\pm }\, ,  
\nn \\
&O_{\cB3++(-;\pm)}^{(2)ab\, \balpha\bt}
= \cB_{n+}^a\, \cB_{\bar n+}^b \, J_{n\bar n\, - }^{\balpha\bt}   J_{e\pm } \, , \qquad 
&&O_{\cB3--(+;\pm)}^{(2)ab\, \balpha\bt}
=    \cB_{n-}^a\, \cB_{\bar n-}^b \, J_{n\bar n\, +}^{\balpha\bt} J_{e\pm }\, , 
\\
&O_{\cB3+-(-;\pm)}^{(2)ab\, \balpha\bt}
=    \cB_{n+}^a\, \cB_{\bar n-}^b \, J_{n\bar n\, - }^{\balpha\bt}  J_{e\pm }\, , \qquad 
&&O_{\cB3-+(+;\pm)}^{(2)ab\, \balpha\bt}
=\cB_{n-}^a\, \cB_{\bar n+}^b  \, J_{n\bar n\, + }^{\balpha\bt} J_{e\pm }  \, , \nn
\end{alignat}
where we have used angular momentum to eliminate operators such as $\cB_{n+}^a\, \cB_{\bar n-}^b \, J_{n\bar n\, + }^{\balpha\bt}   J_{e\pm }$ and $\cB_{n-}^a\, \cB_{\bar n+}^b \, J_{n\bar n\, - }^{\balpha\bt}   J_{e\pm }$. Here the post-BPS color basis is given by
\begin{equation}
\vT_{\BPS}^{\, ab \alpha\bbeta}
= \Bigl(
(T^a  Y^\dagger_n Y_{\bar n} T^b)_{\alpha\bbeta}\,,\, (Y^\dagger_n T^d \cY_{\bn}^{db}  T^c \cY^{ca}_n Y_{\bar n})_{\alpha\bbeta} \,,\, \tr[ T^c \cY^{ca}_n T^d \cY_{\bn}^{db} ]\, [Y^\dagger_n Y_{\bar n}]_{\alpha\bbeta}
\Bigr)
.
\end{equation}
This is easily obtained following the steps described below \Eq{eq:BPS1}.

The next relevant case is when the gluons are in distinct collinear sectors and the quarks are in the same collinear sector. Here, the basis of helicity operators is
\begin{align}
 \boldsymbol{(g q\bar q)_n (g)_{\bn}:}   {\vcenter{\includegraphics[width=0.18\columnwidth]{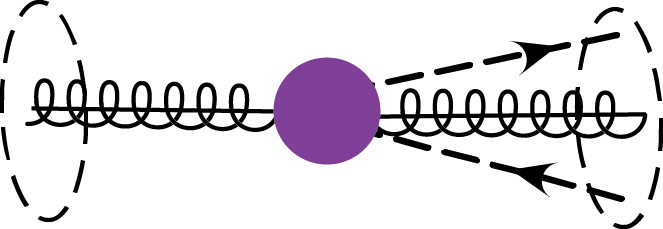}}} \nn
\end{align}
\vspace{-0.4cm}
\begin{alignat}{2} \label{eq:eeqqgg_basis3}
&O_{\cB4++(0:\pm)}^{(2)ab\, \balpha\bt}
=   \cB_{n+}^a\, \cB_{\bar n+}^b \, J_{n\,{0} }^{\balpha\bt}  J_{e\pm }\, , \qquad 
&&O_{\cB4++(\bar 0:\pm)}^{(2)ab\, \balpha\bt}
=\cB_{n+}^a\, \cB_{\bar n+}^b  \, J_{n\,{\bar 0} }^{\balpha\bt}   J_{e\pm }\, ,  \\
&O_{\cB4--(0:\pm)}^{(2)ab\, \balpha\bt}
=  \cB_{n-}^a\, \cB_{\bar n-}^b \, J_{n\,{0} }^{\balpha\bt}    J_{e\pm }\, , \qquad 
&&O_{\cB4--(\bar 0:\pm)}^{(2)ab\, \balpha\bt}
= \cB_{ n-}^a\, \cB_{\bar n-}^b  \, J_{n\,{\bar 0} }^{\balpha\bt}   J_{e\pm }\, . \nn
\end{alignat}
In writing \eq{eeqqgg_basis3} we have again used constraints of angular momentum conservation to restrict the allowed operators in the basis (e.g. we have eliminated $\cB_{n+}^a\, \cB_{\bar n-}^b \, J_{n\,{0} }^{\balpha\bt}  J_{e\pm }$). The color basis after BPS field redefinition in this case is
\begin{equation}
\vT_{\BPS}^{\, ab \alpha\bbeta}
= \Bigl(
(T^a Y^\dagger_n Y_{\bar n} T^b Y^\dagger_{\bar n} Y_n)_{\alpha\bbeta}\,,\, (Y^\dagger_n Y_{\bar n} T^bY^\dagger_{\bar n} Y_n T^a)_{\alpha\bbeta} \,,\, \tr[T^c \cY_{n}^{ca} T^d \cY_{\bn}^{db}]\, \delta_{\alpha\bbeta}
\Bigr)
\,.
\end{equation}

Finally, we consider the basis of operators with both quarks in the same collinear sector, and both gluons in the other collinear sector. Imposing angular momentum conservation reduces the basis from four to two distinct operators
\begin{align}
& \boldsymbol{(q \bar q)_n (gg)_{\bn}:}{\vcenter{\includegraphics[width=0.18\columnwidth]{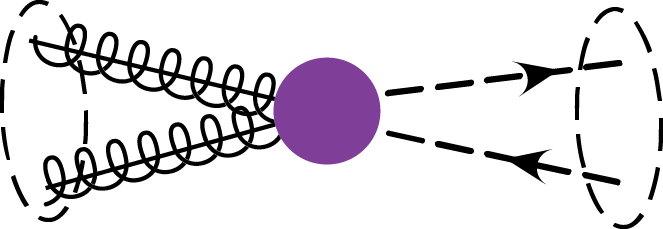}}}  \nn
\end{align}
\vspace{-0.4cm}
\begin{align}\label{eq:eeqqgg_basis4}
&O_{\cB5+-(0:\pm)}^{(2)ab\, \balpha\bt}
= \cB_{\bar n+}^a\, \cB_{ \bar n-}^b  \, J_{n\,{0} }^{\balpha\bt}   J_{e\pm }\,, \qquad 
&O_{\cB5+-(\bar 0:\pm)}^{(2)ab\, \balpha\bt}
= \cB_{\bar n+}^a\, \cB_{ \bar n-}^b  \, J_{n\,{\bar 0} }^{\balpha\bt}   J_{e\pm }\,.
\end{align}
Here, the color basis after BPS field redefinition is
\begin{equation}
\vT_{\BPS}^{\, ab \alpha\bbeta}
= \Bigl(
(Y_n^\dagger Y_\bn T^a T^b Y^\dagger_{\bar n} Y_n )_{\alpha\bbeta}
\,,\, 
(Y_n^\dagger Y_\bn T^b T^a Y^\dagger_{\bar n} Y_n )_{\alpha\bbeta}
\,,\, 
\tr[T^a T^b]\, \delta_{\alpha\bbeta} \Bigr)
\,.
\end{equation}

These operators, provide a complete basis of hard scattering operators with two back to back collinear sectors in the $q\bar qgg$ channel. This example illustrates several key aspects of using the subleading helicity operators: imposing the angular momentum constraints has helped reduce the number of distinct helicity labels that we must consider, the structure of the ultrasoft Wilson lines is determined by the BPS field redefinition and the enumeration of a complete basis is as simple as writing down all allowed helicity choices. The analysis of this channel only gives partial results for the ${\cal O}(\lambda^2)$ operator basis. The full basis of subleading operators for the back-to-back case at $\cO(\lambda)$ and $\cO(\lambda^2)$ will be discussed in detail in~\cite{subhel:long}, including an analysis of relations that occur from parity and charge conjugation.

\section{Conclusions}\label{sec:conclusions}

In this paper we have defined a complete set of helicity operator building blocks which can be used to construct operators at any order in the SCET power expansion, extending the leading power construction of~\cite{Moult:2015aoa}. These building blocks are summarized in \Tab{tab:helicityBB}, and are each collinear and ultrasoft gauge invariant in \SCETi. They include two collinear gluon fields, three ultrasoft gluon fields, two types of derivatives, and various bilinear fermion currents constructed from collinear and ultrasoft fields.  The use of gauge invariant building blocks allows for a simple organization of color structures, and generalizes the color bases familiar from the study of on-shell amplitudes to include the ultrasoft Wilson lines describing the eikonalized particles involved in the scattering process. We also discussed the appearance of interesting angular momentum selection rules which first become nontrivial at subleading power, when multiple fields appear in the same collinear sector. The efficiency of the helicity operator building blocks for constructing minimal bases, as well as the angular momentum selection rules, were demonstrated by constructing an $\mathcal{O}(\lambda^2)$ basis of $q\bar qgg$ operators with two hard scattering directions. These operators are required for the study of $e^+e^-\to$ dijets or Drell-Yan, at subleading power.

A key application of the ideas in this paper is to the calculation of subleading power corrections to physical observables of phenomenological interest. While leading power factorization and resummation has been widely applied to DIS, $e^+e^-\to$ jets and hadron collider observables (see e.g. \cite{Catani:1992ua,Bozzi:2003jy,Dasgupta:2003iq,Banfi:2004nk,Becher:2006mr,Fleming:2007qr,Becher:2007ty,Fleming:2007xt,Becher:2008cf,Stewart:2009yx,Stewart:2010pd,Chien:2010kc,Abbate:2010xh,Feige:2012vc,Becher:2012qa,Banfi:2012jm,Chien:2012ur,Jouttenus:2013hs,Kang:2013nha,Stewart:2013faa,Hoang:2014wka,Larkoski:2015kga} for a non-exhaustive selection), the complexity of subleading factorization has rendered it impractical despite its theoretical and phenomenological importance. In a companion paper~\cite{subhel:long}, we will provide a more detailed discussion of the subleading helicity building blocks introduced here, including a construction of a complete basis of all operators needed for two hard scattering directions up to $\mathcal{O}(\lambda^2)$ (including operators for other partonic channels, $\cP_\perp$ insertions, etc.).  Symmetry arguments, which are manifest in a helicity operator basis, simplify the construction of operators, and also many aspects of their use for factorizing amplitudes and cross sections. We expect that the use of helicity inspired methods will prove useful in the future study of the subleading singular limits of gauge theories, and of factorization theorems at subleading power.

\begin{acknowledgments}
We thank Ilya Feige for collaboration on related work as well as Wouter Waalewijn and Frank Tackmann for helpful discussions. We thank the Erwin Schr\"odinger Institute and the organizers of the ``Jets and Quantum Fields for LHC and Future Colliders'' workshop for hospitality and support while portions of this work were completed. This work was supported in part by the Office of Nuclear Physics of the U.S. Department of Energy under the Grant No.~DE-SC0011090, and by NSERC of Canada. I.S. was also supported by the Simons Foundation through the Investigator grant 327942.

\end{acknowledgments}

\appendix

\bibliography{subhelopsbib}{}

\providecommand{\href}[2]{#2}\begingroup\raggedright\begin{thebibliography}{100}

\bibitem{Dixon:1996wi}
L.~J. Dixon, {\it {Calculating scattering amplitudes efficiently}},
  \href{http://arxiv.org/abs/hep-ph/9601359}{{\tt hep-ph/9601359}}.

\bibitem{Elvang:2013cua}
H.~Elvang and Y.-t. Huang, {\it {Scattering Amplitudes}},
  \href{http://arxiv.org/abs/1308.1697}{{\tt arXiv:1308.1697}}.

\bibitem{Dixon:2013uaa}
L.~J. Dixon, {\it {A brief introduction to modern amplitude methods}},
  \href{http://arxiv.org/abs/1310.5353}{{\tt arXiv:1310.5353}}.

\bibitem{Henn:2014yza}
J.~M. Henn and J.~C. Plefka, {\it {Scattering Amplitudes in Gauge Theories}},
  {\em Lect. Notes Phys.} {\bf 883} (2014) 1--195.

\bibitem{Cheung:2015aba}
C.~Cheung and C.-H. Shen, {\it {Nonrenormalization Theorems without
  Supersymmetry}},  {\em Phys. Rev. Lett.} {\bf 115} (2015), no.~7 071601,
  [\href{http://arxiv.org/abs/1505.01844}{{\tt arXiv:1505.01844}}].

\bibitem{Alonso:2014rga}
R.~Alonso, E.~E. Jenkins, and A.~V. Manohar, {\it {Holomorphy without
  Supersymmetry in the Standard Model Effective Field Theory}},  {\em Phys.
  Lett.} {\bf B739} (2014) 95--98, [\href{http://arxiv.org/abs/1409.0868}{{\tt
  arXiv:1409.0868}}].

\bibitem{Moult:2015aoa}
I.~Moult, I.~W. Stewart, F.~J. Tackmann, and W.~J. Waalewijn, {\it {Employing
  Helicity Amplitudes for Resummation}},
  \href{http://arxiv.org/abs/1508.02397}{{\tt arXiv:1508.02397}}.

\bibitem{Bauer:2000ew}
C.~W. Bauer, S.~Fleming, and M.~E. Luke, {\it {Summing Sudakov logarithms in $B
  \to X_s\gamma$ in effective field theory}},  {\em Phys. Rev. D} {\bf 63}
  (2000) 014006, [\href{http://arxiv.org/abs/hep-ph/0005275}{{\tt
  hep-ph/0005275}}].

\bibitem{Bauer:2000yr}
C.~W. Bauer, S.~Fleming, D.~Pirjol, and I.~W. Stewart, {\it An effective field
  theory for collinear and soft gluons: Heavy to light decays},  {\em Phys.
  Rev. D} {\bf 63} (2001) 114020,
  [\href{http://arxiv.org/abs/hep-ph/0011336}{{\tt hep-ph/0011336}}].

\bibitem{Bauer:2001ct}
C.~W. Bauer and I.~W. Stewart, {\it Invariant operators in collinear effective
  theory},  {\em Phys. Lett. B} {\bf 516} (2001) 134--142,
  [\href{http://arxiv.org/abs/hep-ph/0107001}{{\tt hep-ph/0107001}}].

\bibitem{Bauer:2001yt}
C.~W. Bauer, D.~Pirjol, and I.~W. Stewart, {\it Soft-collinear factorization in
  effective field theory},  {\em Phys. Rev. D} {\bf 65} (2002) 054022,
  [\href{http://arxiv.org/abs/hep-ph/0109045}{{\tt hep-ph/0109045}}].

\bibitem{Weinberg:1965nx}
S.~Weinberg, {\it {Infrared photons and gravitons}},  {\em Phys. Rev.} {\bf
  140} (1965) B516--B524.

\bibitem{Collins:1985ue}
J.~C. Collins, D.~E. Soper, and G.~Sterman, {\it {Factorization for Short
  Distance Hadron - Hadron Scattering}},  {\em Nucl. Phys. B} {\bf 261} (1985)
  104.

\bibitem{Collins:1989gx}
J.~C. Collins, D.~E. Soper, and G.~Sterman, {\it {Factorization of Hard
  Processes in QCD}},  {\em Adv. Ser. Direct. High Energy Phys.} {\bf 5} (1988)
  1--91, [\href{http://arxiv.org/abs/hep-ph/0409313}{{\tt hep-ph/0409313}}].

\bibitem{Collins:1988ig}
J.~C. Collins, D.~E. Soper, and G.~Sterman, {\it Soft gluons and
  factorization},  {\em Nucl. Phys. B} {\bf 308} (1988) 833.

\bibitem{Collins:1984kg}
J.~C. Collins, D.~E. Soper, and G.~Sterman, {\it {Transverse Momentum
  Distribution in Drell-Yan Pair and $W$ and $Z$ Boson Production}},  {\em
  Nucl. Phys. B} {\bf 250} (1985) 199.

\bibitem{Catani:1991kz}
S.~Catani, G.~Turnock, B.~R. Webber, and L.~Trentadue, {\it {Thrust
  distribution in e+ e- annihilation}},  {\em Phys. Lett.} {\bf B263} (1991)
  491--497.

\bibitem{Catani:1992ua}
S.~Catani, L.~Trentadue, G.~Turnock, and B.~R. Webber, {\it Resummation of
  large logarithms in $e^+ e^-$ event shape distributions},  {\em Nucl. Phys.
  B} {\bf 407} (1993) 3--42.

\bibitem{Sterman:1986aj}
G.~Sterman, {\it {Summation of Large Corrections to Short Distance Hadronic
  Cross-Sections}},  {\em Nucl. Phys. B} {\bf 281} (1987) 310.

\bibitem{Catani:1989ne}
S.~Catani and L.~Trentadue, {\it {Resummation of the QCD Perturbative Series
  for Hard Processes}},  {\em Nucl. Phys. B} {\bf 327} (1989) 323.

\bibitem{Low:1958sn}
F.~E. Low, {\it {Bremsstrahlung of very low-energy quanta in elementary
  particle collisions}},  {\em Phys. Rev.} {\bf 110} (1958) 974--977.

\bibitem{Burnett:1967km}
T.~H. Burnett and N.~M. Kroll, {\it {Extension of the low soft photon
  theorem}},  {\em Phys. Rev. Lett.} {\bf 20} (1968) 86.

\bibitem{DelDuca:1990gz}
V.~Del~Duca, {\it {High-energy Bremsstrahlung Theorems for Soft Photons}},
  {\em Nucl. Phys.} {\bf B345} (1990) 369--388.

\bibitem{Casali:2014xpa}
E.~Casali, {\it {Soft sub-leading divergences in Yang-Mills amplitudes}},  {\em
  JHEP} {\bf 08} (2014) 077, [\href{http://arxiv.org/abs/1404.5551}{{\tt
  arXiv:1404.5551}}].

\bibitem{Schwab:2014xua}
B.~U.~W. Schwab and A.~Volovich, {\it {Subleading Soft Theorem in Arbitrary
  Dimensions from Scattering Equations}},  {\em Phys. Rev. Lett.} {\bf 113}
  (2014), no.~10 101601, [\href{http://arxiv.org/abs/1404.7749}{{\tt
  arXiv:1404.7749}}].

\bibitem{Cachazo:2014fwa}
F.~Cachazo and A.~Strominger, {\it {Evidence for a New Soft Graviton Theorem}},
   \href{http://arxiv.org/abs/1404.4091}{{\tt arXiv:1404.4091}}.

\bibitem{Larkoski:2014hta}
A.~J. Larkoski, {\it {Conformal Invariance of the Subleading Soft Theorem in
  Gauge Theory}},  {\em Phys. Rev.} {\bf D90} (2014), no.~8 087701,
  [\href{http://arxiv.org/abs/1405.2346}{{\tt arXiv:1405.2346}}].

\bibitem{He:2014bga}
S.~He, Y.-t. Huang, and C.~Wen, {\it {Loop Corrections to Soft Theorems in
  Gauge Theories and Gravity}},  {\em JHEP} {\bf 12} (2014) 115,
  [\href{http://arxiv.org/abs/1405.1410}{{\tt arXiv:1405.1410}}].

\bibitem{Bern:2014oka}
Z.~Bern, S.~Davies, and J.~Nohle, {\it {On Loop Corrections to Subleading Soft
  Behavior of Gluons and Gravitons}},  {\em Phys. Rev.} {\bf D90} (2014), no.~8
  085015, [\href{http://arxiv.org/abs/1405.1015}{{\tt arXiv:1405.1015}}].

\bibitem{Broedel:2014fsa}
J.~Broedel, M.~de~Leeuw, J.~Plefka, and M.~Rosso, {\it {Constraining subleading
  soft gluon and graviton theorems}},  {\em Phys. Rev.} {\bf D90} (2014), no.~6
  065024, [\href{http://arxiv.org/abs/1406.6574}{{\tt arXiv:1406.6574}}].

\bibitem{Zlotnikov:2014sva}
M.~Zlotnikov, {\it {Sub-sub-leading soft-graviton theorem in arbitrary
  dimension}},  {\em JHEP} {\bf 10} (2014) 148,
  [\href{http://arxiv.org/abs/1407.5936}{{\tt arXiv:1407.5936}}].

\bibitem{Lysov:2014csa}
V.~Lysov, S.~Pasterski, and A.~Strominger, {\it {Low's Subleading Soft Theorem
  as a Symmetry of QED}},  {\em Phys. Rev. Lett.} {\bf 113} (2014), no.~11
  111601, [\href{http://arxiv.org/abs/1407.3814}{{\tt arXiv:1407.3814}}].

\bibitem{White:2014qia}
C.~D. White, {\it {Diagrammatic insights into next-to-soft corrections}},  {\em
  Phys. Lett.} {\bf B737} (2014) 216--222,
  [\href{http://arxiv.org/abs/1406.7184}{{\tt arXiv:1406.7184}}].

\bibitem{Larkoski:2014bxa}
A.~J. Larkoski, D.~Neill, and I.~W. Stewart, {\it {Soft Theorems from Effective
  Field Theory}},  \href{http://arxiv.org/abs/1412.3108}{{\tt
  arXiv:1412.3108}}.

\bibitem{Kapec:2015ena}
D.~Kapec, M.~Pate, and A.~Strominger, {\it {New Symmetries of QED}},
  \href{http://arxiv.org/abs/1506.02906}{{\tt arXiv:1506.02906}}.

\bibitem{Strominger:2015bla}
A.~Strominger, {\it {Magnetic Corrections to the Soft Photon Theorem}},
  \href{http://arxiv.org/abs/1509.00543}{{\tt arXiv:1509.00543}}.

\bibitem{Lee:2004ja}
K.~S. Lee and I.~W. Stewart, {\it {Factorization for power corrections to $B
  \rightarrow X_s \gamma$ and $B \rightarrow X_u \ell \bar \nu$}},  {\em
  Nucl.Phys.} {\bf B721} (2005) 325--406,
  [\href{http://arxiv.org/abs/hep-ph/0409045}{{\tt hep-ph/0409045}}].

\bibitem{Mannel:2004as}
T.~Mannel and F.~J. Tackmann, {\it Shape function effects in {B} $\to$ {X}$_c
  \ell \nu$},  \href{http://arxiv.org/abs/hep-ph/0408273}{{\tt
  hep-ph/0408273}}.

\bibitem{Bosch:2004cb}
S.~W. Bosch, M.~Neubert, and G.~Paz, {\it {Subleading shape functions in
  inclusive B decays}},  {\em JHEP} {\bf 11} (2004) 073,
  [\href{http://arxiv.org/abs/hep-ph/0409115}{{\tt hep-ph/0409115}}].

\bibitem{Beneke:2004in}
M.~Beneke, F.~Campanario, T.~Mannel, and B.~D. Pecjak, {\it {Power corrections
  to $\bar B \to X_u \ell \bar\nu (X_s \gamma)$ decay spectra in the
  'shape-function' region}},  {\em JHEP} {\bf 06} (2005) 071,
  [\href{http://arxiv.org/abs/hep-ph/0411395}{{\tt hep-ph/0411395}}].

\bibitem{Laenen:2008gt}
E.~Laenen, G.~Stavenga, and C.~D. White, {\it {Path integral approach to
  eikonal and next-to-eikonal exponentiation}},  {\em JHEP} {\bf 0903} (2009)
  054, [\href{http://arxiv.org/abs/0811.2067}{{\tt arXiv:0811.2067}}].

\bibitem{Laenen:2008ux}
E.~Laenen, L.~Magnea, and G.~Stavenga, {\it {On next-to-eikonal corrections to
  threshold resummation for the Drell-Yan and DIS cross sections}},  {\em
  Phys.Lett.} {\bf B669} (2008) 173--179,
  [\href{http://arxiv.org/abs/0807.4412}{{\tt arXiv:0807.4412}}].

\bibitem{Grunberg:2009yi}
G.~Grunberg and V.~Ravindran, {\it {On threshold resummation beyond leading 1-x
  order}},  {\em JHEP} {\bf 10} (2009) 055,
  [\href{http://arxiv.org/abs/0902.2702}{{\tt arXiv:0902.2702}}].

\bibitem{Laenen:2010uz}
E.~Laenen, L.~Magnea, G.~Stavenga, and C.~D. White, {\it {Next-to-eikonal
  corrections to soft gluon radiation: a diagrammatic approach}},  {\em JHEP}
  {\bf 1101} (2011) 141, [\href{http://arxiv.org/abs/1010.1860}{{\tt
  arXiv:1010.1860}}].

\bibitem{Almasy:2010wn}
A.~A. Almasy, G.~Soar, and A.~Vogt, {\it {Generalized double-logarithmic
  large-x resummation in inclusive deep-inelastic scattering}},  {\em JHEP}
  {\bf 03} (2011) 030, [\href{http://arxiv.org/abs/1012.3352}{{\tt
  arXiv:1012.3352}}].

\bibitem{Freedman:2013vya}
S.~M. Freedman, {\it {Subleading Corrections To Thrust Using Effective Field
  Theory}},  \href{http://arxiv.org/abs/1303.1558}{{\tt arXiv:1303.1558}}.

\bibitem{Freedman:2014uta}
S.~M. Freedman and R.~Goerke, {\it {Renormalization of Subleading Dijet
  Operators in Soft-Collinear Effective Theory}},
  \href{http://arxiv.org/abs/1408.6240}{{\tt arXiv:1408.6240}}.

\bibitem{Bonocore:2014wua}
D.~Bonocore, E.~Laenen, L.~Magnea, L.~Vernazza, and C.~D. White, {\it {The
  method of regions and next-to-soft corrections in Drell--Yan production}},
  {\em Phys. Lett.} {\bf B742} (2015) 375--382,
  [\href{http://arxiv.org/abs/1410.6406}{{\tt arXiv:1410.6406}}].

\bibitem{deFlorian:2014vta}
D.~de~Florian, J.~Mazzitelli, S.~Moch, and A.~Vogt, {\it {Approximate N$^{3}$LO
  Higgs-boson production cross section using physical-kernel constraints}},
  {\em JHEP} {\bf 10} (2014) 176, [\href{http://arxiv.org/abs/1408.6277}{{\tt
  arXiv:1408.6277}}].

\bibitem{Bonocore:2015esa}
D.~Bonocore, E.~Laenen, L.~Magnea, S.~Melville, L.~Vernazza, and C.~D. White,
  {\it {A factorization approach to next-to-leading-power threshold
  logarithms}},  {\em JHEP} {\bf 06} (2015) 008,
  [\href{http://arxiv.org/abs/1503.05156}{{\tt arXiv:1503.05156}}].

\bibitem{Beneke:2002ni}
M.~Beneke and T.~Feldmann, {\it {Multipole expanded soft collinear effective
  theory with nonAbelian gauge symmetry}},  {\em Phys. Lett.} {\bf B553} (2003)
  267--276, [\href{http://arxiv.org/abs/hep-ph/0211358}{{\tt hep-ph/0211358}}].

\bibitem{Chay:2002vy}
J.~Chay and C.~Kim, {\it {Collinear effective theory at subleading order and
  its application to heavy-light currents}},  {\em Phys. Rev. D} {\bf 65}
  (2002) 114016, [\href{http://arxiv.org/abs/hep-ph/0201197}{{\tt
  hep-ph/0201197}}].

\bibitem{Manohar:2002fd}
A.~V. Manohar, T.~Mehen, D.~Pirjol, and I.~W. Stewart, {\it Reparameterization
  invariance for collinear operators},  {\em Phys. Lett. B} {\bf 539} (2002)
  59--66, [\href{http://arxiv.org/abs/hep-ph/0204229}{{\tt hep-ph/0204229}}].

\bibitem{Pirjol:2002km}
D.~Pirjol and I.~W. Stewart, {\it {A Complete basis for power suppressed
  collinear ultrasoft operators}},  {\em Phys.Rev.} {\bf D67} (2003) 094005,
  [\href{http://arxiv.org/abs/hep-ph/0211251}{{\tt hep-ph/0211251}}].

\bibitem{Beneke:2002ph}
M.~Beneke, A.~Chapovsky, M.~Diehl, and T.~Feldmann, {\it {Soft collinear
  effective theory and heavy to light currents beyond leading power}},  {\em
  Nucl. Phys. B} {\bf 643} (2002) 431--476,
  [\href{http://arxiv.org/abs/hep-ph/0206152}{{\tt hep-ph/0206152}}].

\bibitem{Bauer:2003mga}
C.~W. Bauer, D.~Pirjol, and I.~W. Stewart, {\it {On Power suppressed operators
  and gauge invariance in SCET}},  {\em Phys.Rev.} {\bf D68} (2003) 034021,
  [\href{http://arxiv.org/abs/hep-ph/0303156}{{\tt hep-ph/0303156}}].

\bibitem{Arnesen:2006vb}
C.~M. Arnesen, Z.~Ligeti, I.~Z. Rothstein, and I.~W. Stewart, {\it {Power
  Corrections in Charmless Nonleptonic B-Decays: Annihilation is Factorizable
  and Real}},  {\em Phys. Rev.} {\bf D77} (2008) 054006,
  [\href{http://arxiv.org/abs/hep-ph/0607001}{{\tt hep-ph/0607001}}].

\bibitem{Arnesen:2006dc}
C.~M. Arnesen, I.~Z. Rothstein, and I.~W. Stewart, {\it {Three-parton
  contributions to $B \to M_1 M_2$ annihilation at leading order}},  {\em Phys.
  Lett.} {\bf B647} (2007) 405--412,
  [\href{http://arxiv.org/abs/hep-ph/0611356}{{\tt hep-ph/0611356}}]. [Erratum:
  Phys. Lett.B653,450(2007)].

\bibitem{Lee:2006nr}
C.~Lee and G.~F. Sterman, {\it {Momentum Flow Correlations from Event Shapes:
  Factorized Soft Gluons and Soft-Collinear Effective Theory}},  {\em
  Phys.~Rev.~D} {\bf 75} (2007) 014022,
  [\href{http://arxiv.org/abs/hep-ph/0611061}{{\tt hep-ph/0611061}}].

\bibitem{Benzke:2010js}
M.~Benzke, S.~J. Lee, M.~Neubert, and G.~Paz, {\it {Factorization at Subleading
  Power and Irreducible Uncertainties in $\bar B\to X_s\gamma$ Decay}},  {\em
  JHEP} {\bf 08} (2010) 099, [\href{http://arxiv.org/abs/1003.5012}{{\tt
  arXiv:1003.5012}}].

\bibitem{Mateu:2012nk}
V.~Mateu, I.~W. Stewart, and J.~Thaler, {\it {Power Corrections to Event Shapes
  with Mass-Dependent Operators}},  {\em Phys. Rev.} {\bf D87} (2013) 014025,
  [\href{http://arxiv.org/abs/1209.3781}{{\tt arXiv:1209.3781}}].

\bibitem{subhel:long}
I.~Feige, D.~W. Kolodrubetz, I.~Moult, and I.~W. Stewart, {\it {Helicity
  Operators for Subleading Factorization}},  {\em Forthcoming Publication}.

\bibitem{Bauer:2002nz}
C.~W. Bauer, S.~Fleming, D.~Pirjol, I.~Z. Rothstein, and I.~W. Stewart, {\it
  Hard scattering factorization from effective field theory},  {\em Phys. Rev.
  D} {\bf 66} (2002) 014017, [\href{http://arxiv.org/abs/hep-ph/0202088}{{\tt
  hep-ph/0202088}}].

\bibitem{Marcantonini:2008qn}
C.~Marcantonini and I.~W. Stewart, {\it {Reparameterization Invariant Collinear
  Operators}},  {\em Phys. Rev.} {\bf D79} (2009) 065028,
  [\href{http://arxiv.org/abs/0809.1093}{{\tt arXiv:0809.1093}}].

\bibitem{Chay:2004zn}
J.~Chay, C.~Kim, Y.~G. Kim, and J.-P. Lee, {\it {Soft Wilson lines in
  soft-collinear effective theory}},  {\em Phys. Rev. D} {\bf 71} (2005)
  056001, [\href{http://arxiv.org/abs/hep-ph/0412110}{{\tt hep-ph/0412110}}].

\bibitem{Arnesen:2005nk}
C.~M. Arnesen, J.~Kundu, and I.~W. Stewart, {\it Constraint equations for
  heavy-to-light currents in {SCET}},  {\em Phys. Rev. D} {\bf 72} (2005)
  114002, [\href{http://arxiv.org/abs/hep-ph/0508214}{{\tt hep-ph/0508214}}].

\bibitem{Manohar:1993qn}
A.~V. Manohar and M.~B. Wise, {\it {Inclusive semileptonic B and polarized
  Lambda(b) decays from QCD}},  {\em Phys. Rev.} {\bf D49} (1994) 1310--1329,
  [\href{http://arxiv.org/abs/hep-ph/9308246}{{\tt hep-ph/9308246}}].

\bibitem{Beneke:2000ry}
M.~Beneke, G.~Buchalla, M.~Neubert, and C.~T. Sachrajda, {\it {QCD
  factorization for exclusive, nonleptonic B meson decays: General arguments
  and the case of heavy light final states}},  {\em Nucl. Phys.} {\bf B591}
  (2000) 313--418, [\href{http://arxiv.org/abs/hep-ph/0006124}{{\tt
  hep-ph/0006124}}].

\bibitem{Beneke:2000wa}
M.~Beneke and T.~Feldmann, {\it {Symmetry breaking corrections to heavy to
  light B meson form-factors at large recoil}},  {\em Nucl. Phys.} {\bf B592}
  (2001) 3--34, [\href{http://arxiv.org/abs/hep-ph/0008255}{{\tt
  hep-ph/0008255}}].

\bibitem{Beneke:2001at}
M.~Beneke, T.~Feldmann, and D.~Seidel, {\it {Systematic approach to exclusive B
  ---> V l+ l-, V gamma decays}},  {\em Nucl. Phys.} {\bf B612} (2001) 25--58,
  [\href{http://arxiv.org/abs/hep-ph/0106067}{{\tt hep-ph/0106067}}].

\bibitem{Bauer:2001cu}
C.~W. Bauer, D.~Pirjol, and I.~W. Stewart, {\it {A Proof of factorization for B
  ---> D pi}},  {\em Phys. Rev. Lett.} {\bf 87} (2001) 201806,
  [\href{http://arxiv.org/abs/hep-ph/0107002}{{\tt hep-ph/0107002}}].

\bibitem{Bosch:2001gv}
S.~W. Bosch and G.~Buchalla, {\it {The Radiative decays B ---> V gamma at
  next-to-leading order in QCD}},  {\em Nucl. Phys.} {\bf B621} (2002)
  459--478, [\href{http://arxiv.org/abs/hep-ph/0106081}{{\tt hep-ph/0106081}}].

\bibitem{Bauer:2002aj}
C.~W. Bauer, D.~Pirjol, and I.~W. Stewart, {\it Factorization and endpoint
  singularities in heavy-to-light decays},  {\em Phys. Rev. D} {\bf 67} (2003)
  071502(R), [\href{http://arxiv.org/abs/hep-ph/0211069}{{\tt
  hep-ph/0211069}}].

\bibitem{Beneke:2003pa}
M.~Beneke and T.~Feldmann, {\it {Factorization of heavy to light form-factors
  in soft collinear effective theory}},  {\em Nucl.Phys.} {\bf B685} (2004)
  249--296, [\href{http://arxiv.org/abs/hep-ph/0311335}{{\tt hep-ph/0311335}}].

\bibitem{Beneke:2004dp}
M.~Beneke, T.~Feldmann, and D.~Seidel, {\it {Exclusive radiative and
  electroweak b ---> d and b ---> s penguin decays at NLO}},  {\em Eur. Phys.
  J.} {\bf C41} (2005) 173--188,
  [\href{http://arxiv.org/abs/hep-ph/0412400}{{\tt hep-ph/0412400}}].

\bibitem{Arnesen:2005ez}
M.~C. Arnesen, B.~Grinstein, I.~Z. Rothstein, and I.~W. Stewart, {\it {A
  Precision model independent determination of |V(ub)| from B ---> pi e nu}},
  {\em Phys. Rev. Lett.} {\bf 95} (2005) 071802,
  [\href{http://arxiv.org/abs/hep-ph/0504209}{{\tt hep-ph/0504209}}].

\bibitem{Lee:2005pwa}
K.~S.~M. Lee, Z.~Ligeti, I.~W. Stewart, and F.~J. Tackmann, {\it {Universality
  and m(X) cut effects in B ---> X(s) l+ l-}},  {\em Phys. Rev.} {\bf D74}
  (2006) 011501, [\href{http://arxiv.org/abs/hep-ph/0512191}{{\tt
  hep-ph/0512191}}].

\bibitem{Lee:2006gs}
K.~S.~M. Lee, Z.~Ligeti, I.~W. Stewart, and F.~J. Tackmann, {\it {Extracting
  short distance information from b ---> s l+ l- effectively}},  {\em Phys.
  Rev.} {\bf D75} (2007) 034016,
  [\href{http://arxiv.org/abs/hep-ph/0612156}{{\tt hep-ph/0612156}}].

\bibitem{Buras:1989xd}
A.~J. Buras and P.~H. Weisz, {\it {QCD Nonleading Corrections to Weak Decays in
  Dimensional Regularization and 't Hooft-Veltman Schemes}},  {\em Nucl. Phys.}
  {\bf B333} (1990) 66.

\bibitem{Dugan:1990df}
M.~J. Dugan and B.~Grinstein, {\it {On the vanishing of evanescent operators}},
   {\em Phys. Lett.} {\bf B256} (1991) 239--244.

\bibitem{Herrlich:1994kh}
S.~Herrlich and U.~Nierste, {\it {Evanescent operators, scheme dependences and
  double insertions}},  {\em Nucl. Phys.} {\bf B455} (1995) 39--58,
  [\href{http://arxiv.org/abs/hep-ph/9412375}{{\tt hep-ph/9412375}}].

\bibitem{Bozzi:2003jy}
G.~Bozzi, S.~Catani, D.~de~Florian, and M.~Grazzini, {\it {The $q_T$ spectrum
  of the Higgs boson at the LHC in QCD perturbation theory}},  {\em Phys. Lett.
  B} {\bf 564} (2003) 65--72, [\href{http://arxiv.org/abs/hep-ph/0302104}{{\tt
  hep-ph/0302104}}].

\bibitem{Dasgupta:2003iq}
M.~Dasgupta and G.~P. Salam, {\it {Event shapes in e+ e- annihilation and deep
  inelastic scattering}},  {\em J. Phys.} {\bf G30} (2004) R143,
  [\href{http://arxiv.org/abs/hep-ph/0312283}{{\tt hep-ph/0312283}}].

\bibitem{Banfi:2004nk}
A.~Banfi, G.~P. Salam, and G.~Zanderighi, {\it {Resummed event shapes at hadron
  - hadron colliders}},  {\em JHEP} {\bf 08} (2004) 062,
  [\href{http://arxiv.org/abs/hep-ph/0407287}{{\tt hep-ph/0407287}}].

\bibitem{Becher:2006mr}
T.~Becher, M.~Neubert, and B.~D. Pecjak, {\it {Factorization and momentum-space
  resummation in deep-inelastic scattering}},  {\em JHEP} {\bf 01} (2007) 076,
  [\href{http://arxiv.org/abs/hep-ph/0607228}{{\tt hep-ph/0607228}}].

\bibitem{Fleming:2007qr}
S.~Fleming, A.~H. Hoang, S.~Mantry, and I.~W. Stewart, {\it {Jets from massive
  unstable particles: Top-mass determination}},  {\em Phys. Rev. D} {\bf 77}
  (2008) 074010, [\href{http://arxiv.org/abs/hep-ph/0703207}{{\tt
  hep-ph/0703207}}].

\bibitem{Becher:2007ty}
T.~Becher, M.~Neubert, and G.~Xu, {\it {Dynamical Threshold Enhancement and
  Resummation in Drell-Yan Production}},  {\em JHEP} {\bf 07} (2008) 030,
  [\href{http://arxiv.org/abs/0710.0680}{{\tt arXiv:0710.0680}}].

\bibitem{Fleming:2007xt}
S.~Fleming, A.~H. Hoang, S.~Mantry, and I.~W. Stewart, {\it {Top Jets in the
  Peak Region: Factorization Analysis with NLL Resummation}},  {\em Phys. Rev.
  D} {\bf 77} (2008) 114003, [\href{http://arxiv.org/abs/0711.2079}{{\tt
  arXiv:0711.2079}}].

\bibitem{Becher:2008cf}
T.~Becher and M.~D. Schwartz, {\it {A Precise determination of $\alpha_s$ from
  LEP thrust data using effective field theory}},  {\em JHEP} {\bf 07} (2008)
  034, [\href{http://arxiv.org/abs/0803.0342}{{\tt arXiv:0803.0342}}].

\bibitem{Stewart:2009yx}
I.~W. Stewart, F.~J. Tackmann, and W.~J. Waalewijn, {\it {Factorization at the
  LHC: From PDFs to Initial State Jets}},  {\em Phys. Rev. D} {\bf 81} (2010)
  094035, [\href{http://arxiv.org/abs/0910.0467}{{\tt arXiv:0910.0467}}].

\bibitem{Stewart:2010pd}
I.~W. Stewart, F.~J. Tackmann, and W.~J. Waalewijn, {\it {The Beam Thrust Cross
  Section for Drell-Yan at NNLL Order}},  {\em Phys. Rev. Lett.} {\bf 106}
  (2011) 032001, [\href{http://arxiv.org/abs/1005.4060}{{\tt
  arXiv:1005.4060}}].

\bibitem{Chien:2010kc}
Y.-T. Chien and M.~D. Schwartz, {\it {Resummation of heavy jet mass and
  comparison to LEP data}},  {\em JHEP} {\bf 1008} (2010) 058,
  [\href{http://arxiv.org/abs/1005.1644}{{\tt arXiv:1005.1644}}].

\bibitem{Abbate:2010xh}
R.~Abbate, M.~Fickinger, A.~H. Hoang, V.~Mateu, and I.~W. Stewart, {\it {Thrust
  at $N^3LL$ with Power Corrections and a Precision Global Fit for
  alphas(mZ)}},  {\em Phys. Rev. D} {\bf 83} (2011) 074021,
  [\href{http://arxiv.org/abs/1006.3080}{{\tt arXiv:1006.3080}}].

\bibitem{Feige:2012vc}
I.~Feige, M.~D. Schwartz, I.~W. Stewart, and J.~Thaler, {\it {Precision Jet
  Substructure from Boosted Event Shapes}},  {\em Phys. Rev. Lett.} {\bf 109}
  (2012) 092001, [\href{http://arxiv.org/abs/1204.3898}{{\tt
  arXiv:1204.3898}}].

\bibitem{Becher:2012qa}
T.~Becher and M.~Neubert, {\it {Factorization and NNLL Resummation for Higgs
  Production with a Jet Veto}},  {\em JHEP} {\bf 1207} (2012) 108,
  [\href{http://arxiv.org/abs/1205.3806}{{\tt arXiv:1205.3806}}].

\bibitem{Banfi:2012jm}
A.~Banfi, P.~F. Monni, G.~P. Salam, and G.~Zanderighi, {\it {Higgs and Z-boson
  production with a jet veto}},  \href{http://arxiv.org/abs/1206.4998}{{\tt
  arXiv:1206.4998}}.

\bibitem{Chien:2012ur}
Y.-T. Chien, R.~Kelley, M.~D. Schwartz, and H.~X. Zhu, {\it {Resummation of Jet
  Mass at Hadron Colliders}},  {\em Phys. Rev. D} {\bf 87} (2013) 014010,
  [\href{http://arxiv.org/abs/1208.0010}{{\tt arXiv:1208.0010}}].

\bibitem{Jouttenus:2013hs}
T.~T. Jouttenus, I.~W. Stewart, F.~J. Tackmann, and W.~J. Waalewijn, {\it {Jet
  Mass Spectra in Higgs $+$ One Jet at NNLL}},  {\em Phys.~Rev.~D} {\bf 88}
  (2013) 054031, [\href{http://arxiv.org/abs/1302.0846}{{\tt
  arXiv:1302.0846}}].

\bibitem{Kang:2013nha}
D.~Kang, C.~Lee, and I.~W. Stewart, {\it {Using 1-Jettiness to Measure 2 Jets
  in DIS 3 Ways}},  {\em Phys. Rev.} {\bf D88} (2013) 054004,
  [\href{http://arxiv.org/abs/1303.6952}{{\tt arXiv:1303.6952}}].

\bibitem{Stewart:2013faa}
I.~W. Stewart, F.~J. Tackmann, J.~R. Walsh, and S.~Zuberi, {\it {Jet $p_T$
  Resummation in Higgs Production at $NNLL'+NNLO$}},
  \href{http://arxiv.org/abs/1307.1808}{{\tt arXiv:1307.1808}}.

\bibitem{Hoang:2014wka}
A.~H. Hoang, D.~W. Kolodrubetz, V.~Mateu, and I.~W. Stewart, {\it {C-parameter
  Distribution at N${}^3$LL$^\prime$ including Power Corrections}},
  \href{http://arxiv.org/abs/1411.6633}{{\tt arXiv:1411.6633}}.

\bibitem{Larkoski:2015kga}
A.~J. Larkoski, I.~Moult, and D.~Neill, {\it {Analytic Boosted Boson
  Discrimination}},  \href{http://arxiv.org/abs/1507.03018}{{\tt
  arXiv:1507.03018}}.

\end{thebibliography}\endgroup
\bibliographystyle{jhep}

\end{document}